\documentclass[epj]{svjour}
\usepackage{graphics}
\newcommand{\xbj}{x_{\mbox {\scriptsize Bj}}}
\begin{document}
\title{Deep Electroproduction of Photons and Mesons on the Deuteron}
\author{F. Cano\inst{1} \and B. Pire\inst{2}
}                     
%
%
\institute{DAPNIA/SPhN, CEA--Saclay,
F-91191 Gif-sur-Yvette Cedex, France. \and CPhT,
{\'E}cole Polytechnique, F-91128 Palaiseau Cedex, France; UMR C7644 of CNRS.}
\date{July 17, 2003 }
%
\abstract{
We study deeply virtual Compton scattering and deep 
exclusive meson electroproduction on a deuteron target.
We model the Generalized Quark Distributions in the deuteron by
using the impulse approximation for the lowest Fock-space
state on the light-cone. We study the properties of the resulting GPDs,
and verify that sum rules violations are quite small in the impulse approximation.
Numerical predictions are given for the unpolarized cross
sections and polarization asymmetries for the kinematical regimes relevant for
JLab experiments and for HERMES at HERA. We conclude that the signal of coherent scattering on 
the deuteron is comparable
to the one on the proton  at least for low momentum transfer, providing support to the
feasibility of the experiments. The short distance structure of the deuteron may thus be 
scrutinized in the near future.
} 
\maketitle
\section{Introduction}
\label{intro}
The study of hard exclusive processes, such as deeply Virtual Compton
Scattering
(DVCS)
\begin{equation}
e A \to e' \gamma A'
\end{equation}
and deep exclusive  meson electroproduction (DEMP)
\begin{equation}
e A \to e' M A'
\end{equation}
where A is a hadron (usually a nucleon, here a deuteron), and M a meson 
(usually a $\rho $ or a $\pi$) or a pair of mesons (of relatively small invariant mass)
in the kinematical domain of a large momentum transfer $Q^2$ between the leptons
but a small momentum transfer ($t$) between the hadrons, has been recently demonstrated
to open the possibility of obtaining a quite complete picture of the hadronic
structure. The information which can be accessed through these experiments is encoded by
 the Generalized Parton Distributions, GPDs\cite{GPD,DGPR} (for recent reviews
see\cite{Guichon:1998xv}), which give in particular information on the transverse 
location of quarks in the hadrons\cite{femto}. Recent measurements of the azimuthal
dependence of the beam spin asymmetry in DVCS
\cite{HERMES01,CEBAF} have provided
experimental evidence to support the validity of the formalism of GPDs and the
underlying QCD factorization of short-distance and long-distance dominated
subprocesses.

The theoretical arguments used in deriving factorization
theorems in QCD for the nucleon\cite{fact} target case can be applied to the
deuteron case as well, and therefore one can develop the formalism of GPDs
for the deuteron\cite{BERGER01}. From the theoretical viewpoint, it is the simplest
and best known nuclear system and represents the most appropriate starting
point to investigate hard exclusive processes off nuclei\cite{KIRCHNER02}.
On the other hand, these processes could offer a new source of information about the
partonic degrees of freedom in nuclei, complementary to the existing
one from deep inelastic scattering. Experimentally, deuteron targets
are quite common and as a matter of fact, DVCS experiments are being
planned or carried out at facilities like CEBAF at JLab and HERMES at HERA, 
where some data have already been released {\cite{Airapetian:2003yv}. One 
should of course distinguish the case where the deuteron serves merely as a source 
of slightly bound protons and neutrons from the case where the deuteron acts as a 
single hadron. In the former case, the scattering is incoherent and the deuteron will
break up during the reaction. In the latter case, to which we devote our study, the deuteron
stays intact after the scattering. The fact that this occurs in a non negligible fraction of 
events is not evident to everybody, since it is usual, but uncorrect, to mix the concepts of hard 
and destructive reactions. Indeed, as estimates given below will show, the very nature of 
deep electroproduction in the forward region is that the target is not violently shattered by
the hard probe. The fragile nature of the deuteron thus does not prevent it from staying intact.
This picture should of course be experimentally tested through the comparison of rates for 
coherent and incoherent electroproduction. The need for a deuteron recoil detector is primordial 
in this respect.

The paper is organized as follows: In section 2 we remind the reader of
 the formalism of Generalized  Parton Distributions for  spin-1 targets in general and the 
 deuteron in particular. 
In section 3 we explain in detail the construction of the impulse approximation to 
evaluate the helicity amplitudes and in section 4 we  derive the deuteron GPDs  
from the helicity amplitudes and  study the properties and implications with a 
numerical model. In section 5 we give the useful formulae for calculating the DVCS
 cross section. In section 6 we show our numerical estimates for
the usual observables
 in the DVCS case and comment on the feasibility of experiments. In section 7 we examine
 the electroproduction of mesons. 
Througout the paper we will limit ourselves to the quark sector of the GPDs, which is a 
good approximation provided the Bjorken 
variable $\xbj$ is not too small. This variable is defined as usual as 
$\xbj= \frac{Q^2}{2 P \cdot q}$, i.e., it is given  in
lab frame by  $\xbj= \frac{Q^2}{2 M \nu}$, where M is the
deuteron mass and $\nu $  the virtual photon energy. Gluon effects will be needed for understanding
higher energies experiments. Previous short reports on our results have been presented at 
recent conferences \cite{CP} 

\section{GPDs in the deuteron: definitions and basic properties}

A parametrization of the non-perturbative matrix
elements which determine the amplitudes in DVCS and DEMP on a spin-one
target were given in terms of nine GPDs for the quark
sector\cite{BERGER01}:

\begin{eqnarray}
  \label{vaten}
V_{\lambda'\lambda} &=&
  \int \frac{d \kappa}{2 \pi}\,
  e^{i x \kappa 2 \bar{P}.n}
  \langle P', \lambda' |\,
  \bar{\psi}(-\kappa  n)\, \gamma.n\, \psi(\kappa n)
  \,| P, \lambda \rangle
\nonumber \\
&=& \sum_{i=1,5}
  \epsilon'^{\ast \beta}  V^{(i)}_{\beta \alpha}\,
  \epsilon^{\alpha}\, H_{i}(x,\xi,t) ,
\\
A_{\lambda'\lambda} &=&
  \int \frac{d \kappa}{2 \pi}\,
  e^{i x \kappa 2 \bar{P}.n}
  \langle P', \lambda' |\,
  \bar{\psi}(-\kappa n)\, \gamma.n \gamma_5\, \psi(\kappa n)
  \,| P, \lambda \rangle
 \nonumber \\
&=&  \sum_{i=1,4}
  \epsilon'^{\ast \beta}  A^{(i)}_{\beta \alpha}\,
  \epsilon^{\alpha}\, \tilde{H}_{i}(x,\xi,t) ,
\label{axvaten}
\end{eqnarray}

\noindent
where $|P, \lambda \rangle$ represents a deuteron state of
momentum $P$ and polarization $\lambda$,  $\bar{P}=(P+P')/2$, and $n^\mu$
is a light-like vector with $n^+=0, \vec{n}_\perp=0$. Due to the spin-one
character of the target, there are more GPD's than in the nucleon case,
but at the same time the set of polarization observables which in
principle could be measured is also richer. Not much is known about
these non-perturbative objects which encode the way quarks are confined in
deuterons, except a limited set of sum rules and some limiting case values. Sum
rules\cite{BERGER01} relate these GPDs to the usual deuteron form factors :
\begin{eqnarray}
\int_{-1}^1 dx H_i(x,\xi,t)&  = & G_i(t) \hspace{3em} (i=1,2,3) ,
\nonumber \\
\int_{-1}^1 dx \tilde{H}_i(x,\xi,t) & = &  \tilde{G}_i(t) \hspace{3em}
(i=1,2) ,
\label{sumr}
\end{eqnarray}
or lead to a null average :
\begin{eqnarray}
\int_{-1}^1 dx H_4(x,\xi,t) &=& \int_{-1}^1 dx \tilde{H}_3(x,\xi,t)
\;=\; 0 ,
\nonumber \\
\int_{-1}^1 dx H_5(x,\xi,t) &=& \int_{-1}^1 dx \tilde{H}_4(x,\xi,t)
\;=\; 0 .
\end{eqnarray}

Taking the forward limit of the matrix elements defining GPDs leads to the
relations
\cite{BERGER01} between GPDs and parton densities in the deuteron (with obvious
notations) as :
\begin{eqnarray}
\label{forward}
H_1(x,0,0)         & = &  \frac{q^1(x) + q^{-1}(x) + q^0(x)}{3} ,
\nonumber \\
H_5(x,0,0)         & = &  q^0(x) - \frac{q^{1}(x) + q^{-1}(x)}{2} ,
\nonumber \\
\tilde{H}_1(x,0,0) & = &  q^1_\uparrow (x)  - q^{-1}_\uparrow (x)
 \rule{0pt}{3ex}
\end{eqnarray}
for $x>0$. The corresponding relations for $x<0$ involve the antiquark
distributions at $-x$, with an overall minus sign in the expressions
for $H_1$ and $H_5$.

In all this paper we will restrict to the quark contribution. Gluon
contributions are expected  to be small at medium energies but should be included
in a more complete description of the process. Therefore we will limit ourselves to values 
of $\xbj$ not smaller than 0.1.

\section{Helicity Amplitudes in the Impulse Approximation}

The impulse approximation is the zeroth order to
explain the photon-nucleus interaction, but it is the first model one
has to analyze since the bulk of the physics is already contained in
it.

\begin{figure}
\resizebox{0.50\textwidth}{!}{%
  \includegraphics{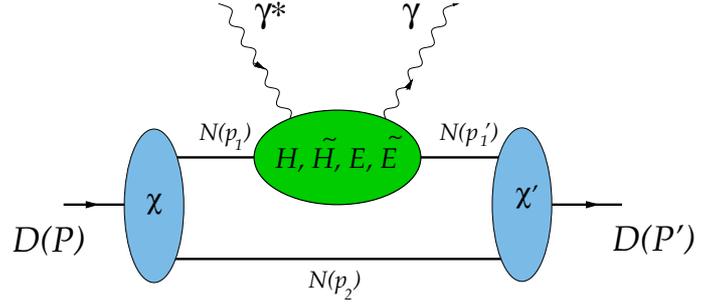}
}
\caption{Estimating the heliticy amplitudes for the $\gamma^* D
  \rightarrow \gamma D$ in the impulse approximation. The final
  result is a convolution model between the deuteron wave function and
  the GPDs for the nucleon (upper blob).}
\end{figure}

In the previous section we mentioned that the relevant quantities are the deuteron GPDs .
 For the sake of simplicity we will model the matrix elements
$V_{\lambda '\lambda}$ (\ref{vaten})and $A_{\lambda ' \lambda}$ (\ref{axvaten}) and recover the GPDs just 
by using the relations given in the appendix. Furthermore, since we are going to limit 
ourselves to the quark content of the deuteron and it is an isoscalar target, we will denote:
\begin{equation}
V_{\lambda ' \lambda}^u = V_{\lambda ' \lambda}^d \equiv V_{\lambda ' \lambda}^q 
\end{equation}

\noindent and a similar relation holds for $A_{\lambda ' \lambda}$.

Let us denote by $P^\mu$ ($P'^\mu$) the momentum of the incoming (outgoing) deuteron 
and $\lambda$ ($\lambda '$) its polarization state, that sometimes we will denote by $0, +$ 
or $-$. To perform this analysis we will choose a symmetric frame where the average 
momentum $\bar{P}^\mu = (P^\mu + P'^\mu)/2$ has no transverse components. We will 
also need a light-like vector $n^\mu$ to define, together with $\bar{P}^\mu$, the light-cone
 and satisfying $\bar{P} \cdot n=1$. To be more concrete, it is convenient to choose a 
 frame where $\bar{P}^\mu$ moves fast to the right.

The momentum transfer $\Delta^\mu = P'^\mu - P^\mu$ has a longitudinal and a transverse
 component. The skewness controls the fraction of momentum transfered in the '+' direction:

\begin{equation}
\xi=-\frac{\Delta \cdot n}{(P + P')\cdot n} = - \frac{\Delta^+}{2 \bar{P}^+}
\end{equation} 

With these considerations, the four-vectors corresponding to each
deuteron are \footnote {we use the following
notation for the four-vectors $(v^+,v^-,\mathbf{v_\perp})$, with
  $v^\pm=\frac{1}{\sqrt{2}}(v^0 \pm v^3)$}:

\begin{eqnarray}
P^\mu & = & ((1+\xi) \bar{P}^+, \frac{M^2+\mathbf{\Delta_\perp}^2/2}{2
  \bar{P}^+ (1+\xi)}, -\frac{\mathbf{\Delta_\perp}}{2}) \\
P'^\mu & = & ((1-\xi) \bar{P}^+, \frac{M^2+\mathbf{\Delta_\perp}^2/2}{2
  \bar{P}^+ (1-\xi)}, \frac{\mathbf{\Delta_\perp}}{2}) 
\end{eqnarray}

The invariant momentum transfer is written as~:

\begin{equation}
t = \Delta^2 = - \frac{4 \xi^2 M^2 +\mathbf{\Delta_\perp}^2 }{1 - \xi^2}
\label{xitrelation}
\end{equation}

The positivity of $\mathbf{\Delta_\perp^2}$ implies that there is a
minimal momentum transfer $t_0$ for a fixed $\xi$:

\begin{equation}
t_0 = - \frac{4 \xi^2 M^2}{1-\xi^2}
\end{equation}

\noindent and at the same time, for a given $t$, there is un upper
bound on the allowed values for $\xi$~:

\begin{equation}
 \xi^2 \le \frac{-t}{4 M^2 - t}
\end{equation}

The GPDs depend in addition of one more variable $x$ which is defined
as the fraction of average momentum carried by the partons in the '+'
direction~:

\begin{equation}
x = \frac{\bar{k} \cdot n}{\bar{P} \cdot n}
\end{equation}

\noindent with $\bar{k}^\mu=(k^\mu + k'^\mu )/2$. Therefore, the
longitudinal momentum of the initial parton is $(x+\xi) \bar{P}^+$,
whereas the final one has $(x-\xi) \bar{P}^+$,  delivering a longitudinal
transfer $\Delta^+=-2 \xi \bar{P}^+$ to the deuteron.

Now let us turn our attention to the kinematics at the nucleon level
and let us define the fraction of longitudinal momentum carried by
each nucleon in the deuteron as~:

\begin{equation}
\alpha_i = \frac{p_i^+}{P^+}~; \; \; \alpha'_i = \frac{p'^+_i}{P'^+}~
\end{equation} 

Therefore, we have that the relevant kinematical quantities for the
nucleons that make up the initial deuteron are ($\alpha \equiv \alpha_1$)~:

\begin{eqnarray}
p_1^+ & = & \alpha (1+ \xi) \bar{P}^+ \nonumber \\
p_2^+ & = & (1-\alpha) (1+ \xi) \bar{P}^+ \\
{\mathbf{p_{1 \perp}}} +{\mathbf{p_{2 \perp}}} &= &  -
\frac{\mathbf{\Delta_\perp}}{2} \nonumber 
\end{eqnarray}

\noindent and for the final deuteron we have($\alpha' \equiv \alpha'_1$)~:

\begin{eqnarray}
p_1'^+ & = & \alpha' (1- \xi) \bar{P}^+ \nonumber \\
p_2'^+ & = & (1-\alpha') (1- \xi) \bar{P}^+ \\
{\mathbf{p'_{1 \perp}}} +{\mathbf{p'_{2 \perp}}} &= &  
\frac{\mathbf{\Delta_\perp}}{2} \nonumber 
\end{eqnarray}

We can now use the decomposition of the deuteron states in terms of
nucleon states and the wave function defined in Appendix A1, Eq. (\ref{deuteronstate}), 
to get~:

\begin{eqnarray}
V^q_{\lambda ' \lambda} & =&   \frac{2}{(16\pi^3)}\int \; d\alpha \,
d\alpha '\,  d\mathbf{p_{1_ \perp}}  d\mathbf{p'_{1_ \perp}}  \, 
\sqrt{\frac{1+\xi}{1-\xi}}
\frac{1}{\sqrt{\alpha \alpha '}} \nonumber \\ & \cdot &
\delta^2 (\mathbf{p'_{1 \perp}} - \mathbf{p_{1 \perp}}-
\mathbf{\Delta_\perp}) 
\delta (\alpha ' - \frac{\alpha (1+\xi) - 2 \xi
}{1-\xi}) \nonumber \\
& \cdot & \Theta (\alpha (1+\xi) - |x| - \xi) ~  \Theta (\alpha (1+\xi) - 2 \xi)  \\ 
& \cdot &
\sum_{\lambda_1 ',\lambda_1,\lambda_2} \chi^*_{\lambda
'}(\alpha',\mathbf{k'_\perp},\lambda_1 ',\lambda_2)
\chi_{\lambda}(\alpha,\mathbf{k}_\perp,\lambda_1,\lambda_2)
\nonumber \\ & \cdot & 
\frac{1}{2}\int \frac{d \kappa}{2 \pi} e^{i \kappa x} \langle p_1', \lambda_1' |
\bar\psi_q(-\frac{\kappa}{2} n) \gamma \cdot n \psi_q(\frac{\kappa}{2} n)
| p_1, \lambda_1 \rangle \nonumber
\label{vintermediate}
\end{eqnarray}

In the equation above, the variables of the spectator nucleon have
been eliminated just by using the normalization properties of the
one-particle states. The arguments of the deuteron wave function (see
appendix A1 for details), $\mathbf{k'_\perp}$ and $\mathbf{k_\perp}$, are the
transverse momentum of the active nucleon in a frame where
$\mathbf{P'_\perp}=0$ and  $\mathbf{P_\perp}=0$ respectively. Their
relationship with the transverse momentum in the symmetric frame is~:

\begin{eqnarray}
\mathbf{k_\perp} \equiv \mathbf{k_{1 _\perp}} & = & \mathbf{p_{1 _\perp}} + \alpha
\frac{\mathbf{\Delta_\perp}}{2} \\ 
\mathbf{k'_\perp} \equiv \mathbf{k'_{1 _\perp}} & = & \mathbf{p'_{1 _\perp}} - \alpha'
\frac{\mathbf{\Delta_\perp}}{2}
\end{eqnarray}

The Heaviside functions in Eq. (\ref{vintermediate}) ensure the
positivity of the 'plus' momentum carried by the nucleons and put a
lower bound on the integration over $\alpha$. The first
one stands for processes where $|x|> \xi$ whereas the second one acts
when we are on the ERBL region where  $|x|< \xi$. 

Therefore, we  end up with matrix elements of a  non-local quark 
operator between one-nucleon states, which is parameterized in terms
of nucleon GPDs. To take advantage of usual parameterizations found in
the literature, it is convenient to keep on working on a symmetric
frame. However, since nucleons  carry some transverse momentum within
the deuteron, the symmetric frame for the deuteron is not the
symmetric frame for the active nucleon. 

By performing a transverse boost it is possible to evaluate the matrix
element in (\ref{vintermediate}) in a symmetric frame, which has the property:

\begin{equation}
\mathbf{\tilde{p}'_{1_\perp}} + \mathbf{\tilde{p}_{1_\perp}} =0
\end{equation}

\noindent where we have marked with a $\tilde{}$ ~ the quantities in
this boosted frame. Since it is a transverse boost, it does not change
the '+' components of the vectors. The light-like vector $n^\mu$ is
not changed either, since it is a light-like vector with no transverse
components, i.e. $\tilde{n}^\mu = n^\mu$. 

In that frame, the parameterization of the nucleon GPDs is made with
variables that refer to the initial and final nucleon, i.e., we
define~:

\begin{eqnarray}
x_N & = & \frac{\bar{\tilde{k}} \cdot \tilde{n}}
{\bar{\tilde{p}}_1 \cdot \tilde{n}} = \frac{x}{\alpha (1+\xi)-\xi} \\
\xi_N & = & - \frac{\tilde{\Delta} \cdot \tilde{n}}
{2 \bar{\tilde{p}}_1 \cdot \tilde{n}} = \frac{\xi}{\alpha (1+\xi)-\xi} 
\end{eqnarray}  

Due to the lower bound on the values of $\alpha$, we have  $\xi_N
\geq \xi$. Moreover, it can be checked that $\frac{x}{\xi} =
\frac{x_N}{\xi_N}$, which is consistent with the fact that we are probing
$q\bar{q}$ distribution amplitudes in the deuteron only through the
nucleon. In other words, when we enter the ERBL region in the
deuteron (i.e. when  $|x|= \xi$), we do so at the nucleon level (i.e.
 when  $|x_{N}«|= \xi_{N}«$).   

The transverse momentum of the nucleon that interacts with the photon
is, after the boost~:

\begin{eqnarray}
\mathbf{\tilde{p}_{1_\perp}} & = &  - \frac
{\mathbf{\tilde{\Delta}_\perp}}{2} \\ 
\mathbf{\tilde{p}'_{1_\perp}} & = &   \frac
{\mathbf{\tilde{\Delta}_\perp}}{2} \\
\mathbf{\tilde{\Delta}_{\perp}} & = &  (1+\xi_N) \mathbf{\Delta_\perp}
+ 2 \xi_N \mathbf{p_{1_\perp}}
\end{eqnarray}

Now we can use the parameterization of the
nucleon matrix element given in the appendix, and after some algebra
and changes in the integration variables we reach the final result for
$V_{\lambda ` \lambda}$~:

\begin{eqnarray}
V^q_{\lambda ' \lambda} & =  & \frac{2}{(16\pi^3)}\int_{\alpha_{min}} \; d\alpha \,
d\alpha '\,  d\mathbf{k}_\perp d\mathbf{k}_\perp ' \, \sqrt{\frac{1+\xi}{1-\xi}}
\frac{1}{\sqrt{\alpha \alpha '}} \nonumber \\ & \cdot &
\delta^2 (\mathbf{k}_\perp ' - \mathbf{k}_\perp - \left( \frac{1-\alpha}{1-\xi }
\right) \mathbf{\Delta}_\perp) \delta (\alpha ' - \frac{\alpha (1+\xi) - 2 \xi
}{1-\xi})
\nonumber \\ & \cdot &
\sum_{\lambda_1 ',\lambda_1,\lambda_2} \chi^*_{\lambda
'}(\alpha',\mathbf{k}_\perp ',\lambda_1 ',\lambda_2)
\chi_{\lambda}(\alpha,\mathbf{k}_\perp,\lambda_1,\lambda_2)
\nonumber \\ & \cdot & 
\left[  (\sqrt{1-\xi_N^2} H^{\mbox {\scriptsize
IS}}(x_N,\xi_N,t) \right. \nonumber \\ & - &
\frac{\xi_N^2}{\sqrt{1-\xi_N^2}} E^{\mbox {\scriptsize IS}}
(x_N,\xi_N,t))
\delta_{\lambda_1 ' \lambda_1}  
\nonumber \\
& + & \left.  \frac{\sqrt{t_0 - t}}{2 m} \eta_{\lambda_1} E^{\mbox
{\scriptsize IS}}(x_N,\xi_N,t)\delta_{\lambda_1 ',-\lambda_1} \right]
\label{vimpulsefinal}
\end{eqnarray}

and a similar expression for $A^q_{\lambda ' \lambda}$

\begin{eqnarray}
A^q_{\lambda ' \lambda} & =  & \frac{2}{(16\pi^3)}\int_{\alpha_{min}} \; d\alpha \,
d\alpha '\,  d\mathbf{{k}_\perp} d\mathbf{{k}_\perp} ' \, \sqrt{\frac{1+\xi}{1-\xi}}
\frac{1}{\sqrt{\alpha \alpha '}} \nonumber \\ & \cdot &
\delta^2 (\mathbf{k}_\perp ' - \mathbf{k}_\perp - \left( \frac{1-\alpha}{1-\xi }
\right) \mathbf{\Delta}_\perp) \delta (\alpha ' - \frac{\alpha (1+\xi) - 2 \xi
}{1-\xi})
\nonumber \\ & \cdot &
\sum_{\lambda_1 ',\lambda_1,\lambda_2} \chi^*_{\lambda
'}(\alpha',\mathbf{k}_\perp ',\lambda_1 ',\lambda_2)
\chi_{\lambda}(\alpha,\mathbf{k}_\perp,\lambda_1,\lambda_2)
\nonumber \\ & \cdot & \left[ 2 \lambda_1 (\sqrt{1-\xi_N^2}
\tilde{H}^{\mbox {\scriptsize IS}}(x_N,\xi_N,t) \right.
 \nonumber \\ & - &
 \frac{\xi_N^2}{\sqrt{1-\xi_N^2}} \tilde{E}^{\mbox {\scriptsize
IS}}(x_N,\xi_N,t)) \delta_{\lambda_1 ' \lambda_1}  \nonumber \\
& + & \left. 2 \lambda_1 \xi_N \frac{\sqrt{t_0 - t}}{2 m} \eta_{\lambda_1}
\tilde{E}^{\mbox {\scriptsize IS}}(x_N,\xi_N,t)\delta_{\lambda_1
',-\lambda_1} \right]
\label{aimpulsefinal}
\end{eqnarray}

The factor 2 in front of the formulae above stands for the
number of nucleons, so that the isoscalar nucleon GPDs ($H^{\mbox
  {\scriptsize IS}}$, $E^{\mbox {\scriptsize IS}}$, \ldots ) 
is the isoscalar combination within one single nucleon~:

\begin{equation}
H^{\mbox {\scriptsize IS}}(x_N,\xi_N,t)=\frac{1}{2} \left[H^{u}(x_N,\xi_N,t)+
H^{d}(x_N,\xi_N,t)\right].
\end{equation}

The phase that goes with the nucleon helicity flip GPDs is given by~: 

\begin{equation}
\eta_{\lambda}= \frac{2 \lambda \tilde{\Delta}_x - i
\tilde{\Delta}_y}{|\tilde\Delta_\perp|}
\end{equation}

For the sake of clarity we have omitted the Heaviside function in the
integrals above but recall that there is a lower bound 
on the value of $\alpha$, which is

\begin{equation}
\alpha_{min}={\mbox max}\left\{ \frac{2 \xi}{1+\xi} ,
\frac{|x|+\xi}{1+\xi}\right\}
\end{equation}

\section{Deuterons GPDs}

Once we have obtained the helicity amplitudes it is straightforward to
get from them the deuteron GPDs, just from the definitions given in
Eq.(\ref{vaten},\ref{axvaten}). To do so in a simple way, one can use the light-cone
polarization vectors given in \cite{BERGER01}. The analytical
expressions are summarized in the appendix. 

At this point one may argue that definitions of deuteron GPDs were not
actually neccessary to reach the results of the preceding section and
that one can derive the cross sections and observables, directly from  
Eqs. (\ref{vimpulsefinal}) and (\ref{aimpulsefinal}). But it should be
emphasized that the genuine objects that parametrize the hadronic
structure are the GPDs, the rest being just kinematics. The GPDs have
well defined properties in some limits and their analysis could help
us in testing the soundness of a model, as a complementary check to
the comparison with experimental data

\subsection{Deuteron wave function}¥
We  need a specific model for the spatial deuteron wave
function. As can be seen in the appendix, in the lower Fock-space
approximation one can link the light-cone wave function to the
usual instant-form wave function through a identification of
variables. We
have chosen a  parametrization of the spatial wave function given by 
the Paris Potential \cite{LACOMBE81} which has a S-wave supplemented 
with a D-wave  component with a probability of 5.8 \%. We do not expect a strong
dependence on the chosen parametrization for the deuteron wave
function. Most of them are identical in the low-momentum region since
they are strongly constrained by the well known form factors. Differences
between parametrizations are significant only in the large momentum
region. Since we are going to limit ourselves to the low-momentum
transfer region, we will not be especially sensitive to the tail of
the wave function.

Nonetheless, before going through the details of the results, let us discuss some features
of the deuteron GPDs that may be expected from quite general grounds.
The skewness parameter $\xi$ determines the  momentum transfer in the longitudinal 
direction:

\begin{equation}
\Delta^+ \equiv (P'^+-P^+) = -2 \xi \bar{P}^+  \;\; ,
\end{equation}

\noindent and in the generalized Bjorken limit this is entirely fixed
by the kinematics of the virtual photon ($\xi\approx \xbj/2$). In
the impulse approximation, this momentum transfer has to be provided
by the active nucleon, and after that, 
the final state of this active nucleon still
has to fit into the final deuteron. Since the deuteron is a loosely
bound system,  one cannot have a very asymmetrical
sharing of longitudinal momentum between the nucleons and one may thus guess that
the formation of the coherent final state will be strongly suppressed
in the impulse approximation for large skewness.

\begin{figure}
\resizebox{0.50\textwidth}{!}{%
\includegraphics{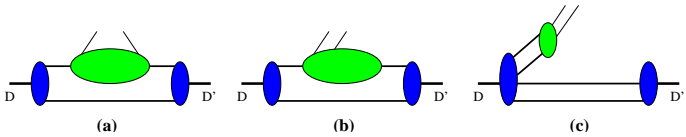}
}
\caption{ Deuteron generalized parton distributions in the impulse (a and b) and 
beyond (c) the impulse approximation}
\label{impulsescheme}
\end{figure}

To be more quantitative let us define the longitudinal momentum distribution of the nucleon 
in the deuteron as:

\begin{equation}
\label{nalpha}
n_\lambda ( \alpha) = \sum_{\lambda_1, \lambda_2} \int
\frac{d\vec{k}_\perp d\beta}{(16 \pi)^3} \; | \chi_\lambda (\beta,
\vec{k}_\perp,\lambda_1,\lambda_2)|^2 \delta (\alpha-\beta) \;\; ,
\end{equation}

\noindent which is normalized according to 

\begin{equation}
\int d\alpha \; n_\lambda(\alpha) =1 \;\; .
\end{equation}

In Fig. \ref{deuteronwidth}.a we show $n_0(\alpha)$ evaluated with the wave function 
from the Paris potential \cite{LACOMBE81}. This distribution is strongly peaked at
 $\alpha=0.5$ and its width is of the order of the ratio of the binding energy divided by the 
 nucleon mass. 

 In the impulse approximation, the active nucleon after the interaction with the photon 
 carries a fraction of longitudinal momentum which is given by

\begin{equation}
\alpha ' = \alpha - \frac{\xbj}{1-\xbj} (1 - \alpha) \;\; . 
\end{equation}

In Fig. \ref{deuteronwidth}.b we plot the difference 
$\alpha - \alpha'$ as a function of
$\alpha$ and for several values of the skewness. We see that for $\xbj
> 0.1$ this difference is larger than the width of the momentum
distribution, and therefore, we will inevitably have  
a too fast or too slow nucleon (in
the longitudinal direction). In this case the central region of
momentum, where a maximal contribution is expected, is missed and then
the 
cross sections will decrease very fast with $\xbj$. In other words,
there is an increasing difficulty in forming a coherent final
state as the longitudinal momentum transfer, i.e. $\xbj$ increases. In
that case other coherent mechanisms, which could involve higher
Fock-space components, will presumably become dominant.
 Not much is known about these states, but it should be
emphasized that the suppression of the diagram of Fig. 1 occurs at
$\xbj$ as low as 0.2, so that there is room to check the importance of
the contribution of these 'exotic' states.

\begin{figure}[t]
\resizebox{0.50\textwidth}{!}{%
\begin{tabular}{lr}
\includegraphics{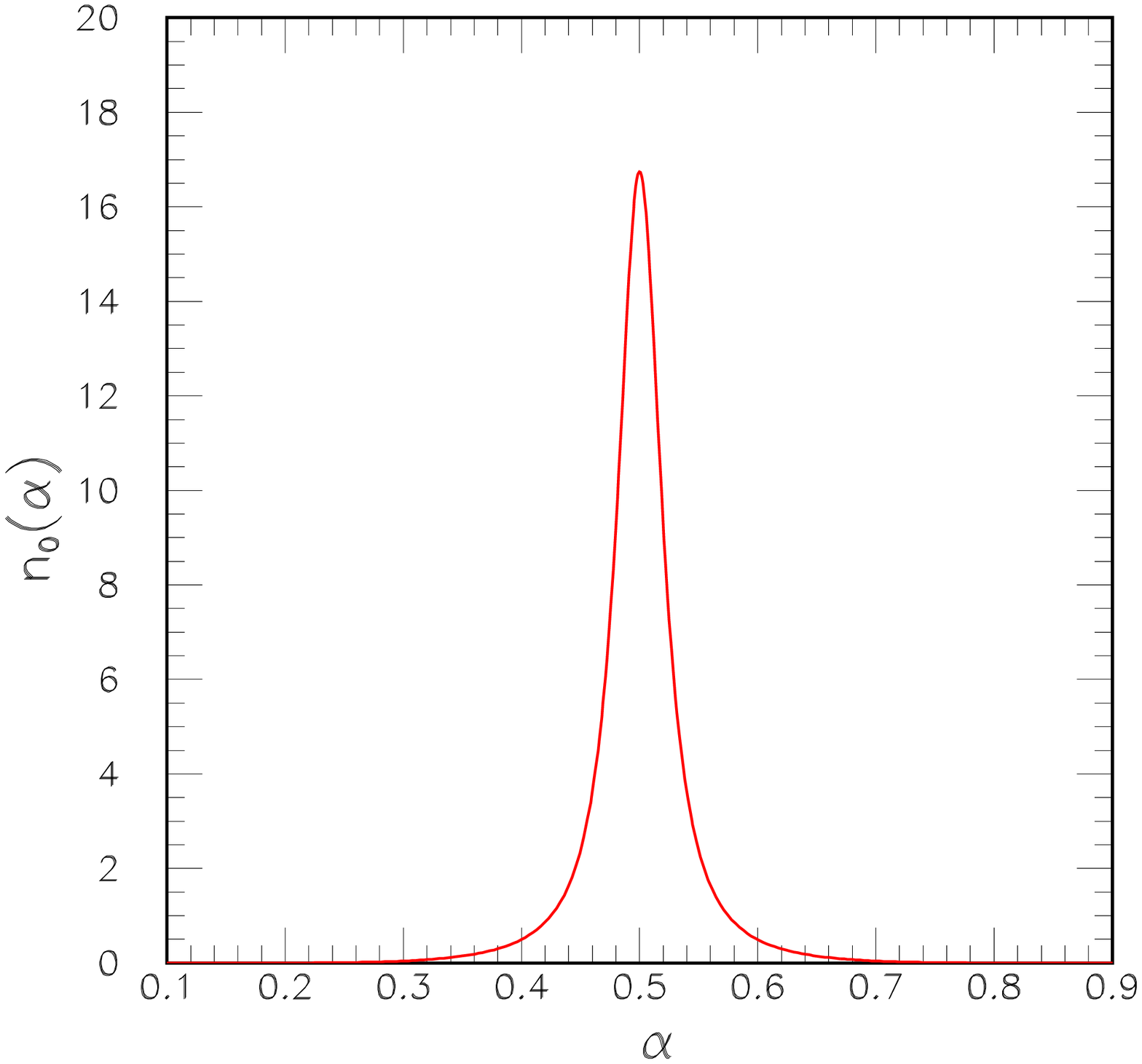} &
\includegraphics{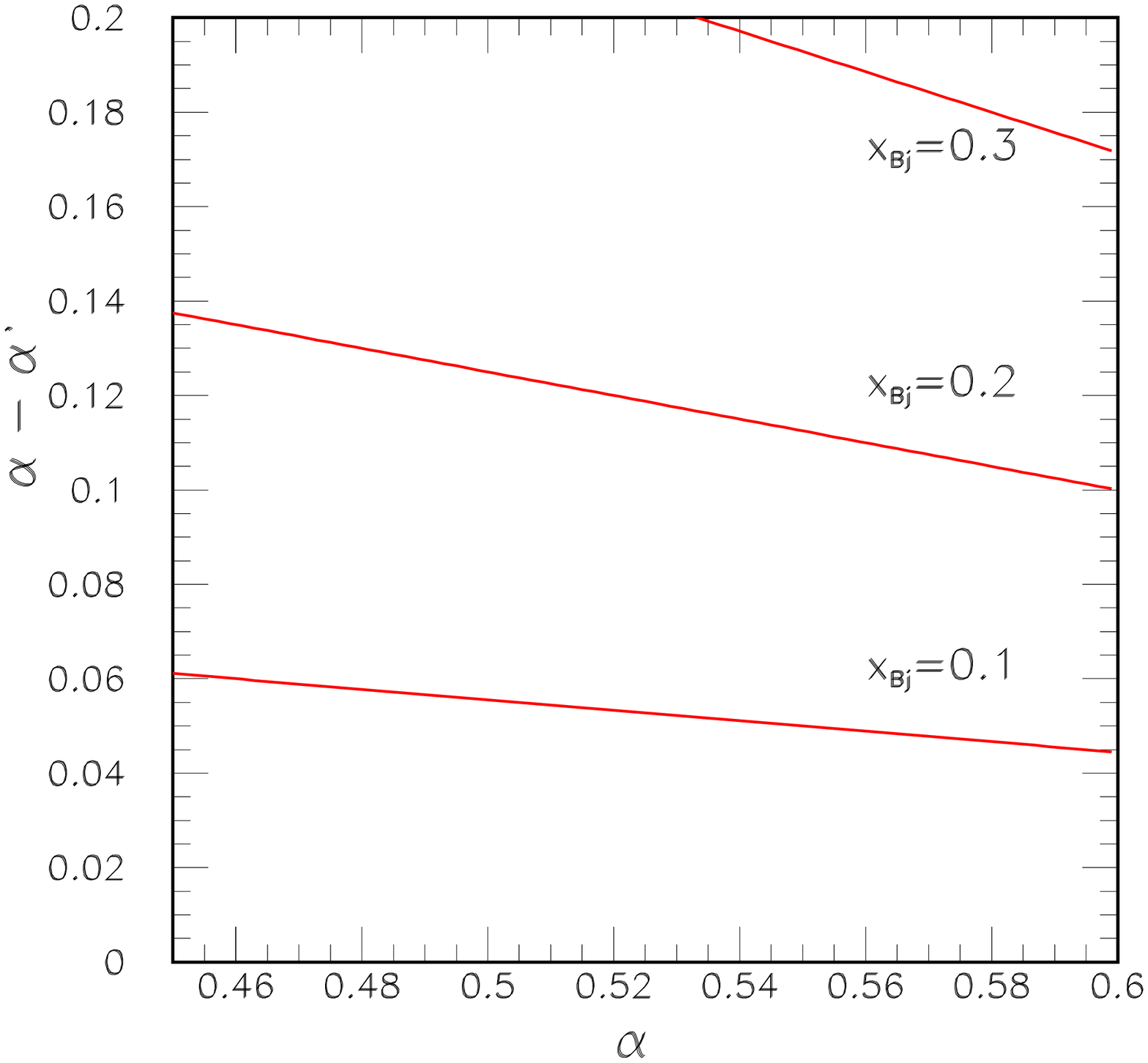}
\end{tabular}
}
\vskip -0.5cm
\caption{(a) Longitudinal momentum distribution of the nucleon within
  the deuteron. (b) Gap between the fractions of longitudinal momentum
  carried by the active nucleon before and after the interaction as a
  function of $\xi (\xbj)$ and $\alpha$}
\label{deuteronwidth}
\end{figure}

The choice of the nucleon GPDs deserves a more detailed discussion.

\subsection{Modelling nucleon GPDs}

Let us first consider the helicity conserving nucleon GPDs. Following
\cite{GOEKE01} we have taken a factorized ansatz for the $t$ dependence of
the nucleon GPDs:

\begin{eqnarray}
H^u(x_N,\xi_N,t) & = & h^u(x_N,\xi_N) \frac{1}{2} F_1^u(t) \\
H^d(x_N,\xi_N,t) & = & h^d(x_N,\xi_N) F_1^d(t) \\
\tilde{H}^q(x_N,\xi_N,t) & = & \tilde{h}^q(x_N,\xi_N) \tilde{F}^q(t)
\end{eqnarray}

\noindent
and neglected the strange quark contributions $H^s$. The flavour
decomposition of the proton and neutron Dirac form factor, for which we have
taken the usual dipole parameterizations \cite{BOFFI93}, gives:

\begin{eqnarray}
F_1^u(t) & = & 2 F_1^p + F_1^n \\
F_1^d(t) & = & 2 F_1^n + F_1^p
\end{eqnarray}

For the axial form factor we have taken $\tilde{F}^q(t) = (1 - t/M_A^2)^{-2}$
with $M_A=1.06$ GeV \cite{AHRENS87}. For  $h^q$ and  $\tilde{h}^q$ we follow
the ansatz based on double distributions:

\begin{eqnarray}
h^q(x_N,\xi_N) &=& \int_{0}^{1} dx'
\int_{-1+x'}^{1-x'} dy' \,
\Bigg[
\delta(x_N-x'- \xi_N y') \, q(x')
\nonumber\\
&&  {}-
\delta(x_N+x' - \xi_N y') \, \bar{q}(x') \,
\Bigg] \, \pi(x',y') ,
\label{ddmodel-h}
\\
\tilde{h}^q(x_N,\xi_N)&=&\int_{0}^{1} dx'
\int_{-1+x'}^{1-x'} dy' \,
\delta(x_N-x' - \xi_N y')\,\nonumber
\\
&& \cdot \Delta q_V(x')\, \pi(x',y') ,
\label{ddmodel-ht}
\\
\pi(x',y') &=& \frac{3}{4}\,
\frac{(1-x')^2 - y'^2}{(1-x')^3} .
\label{profile}
\end{eqnarray}

\noindent where we have only considered the polarization of the valence
quarks. To avoid  numerical problems with the integrals in the low $x«$
region we have followed the procedure explained in \cite{BERGER01}.
Throughout this work we have taken the parameterization MRST 2001 NLO
\cite{MRST01} for the unpolarized parton distributions and the
parameterization LSS01 \cite{LSS01} for the polarized ones.

Concerning the helicity flip nucleon GPDs, $E^q$ and $\tilde{E}^q$, we can
safely neglect the latter since we deal with an isoscalar target. The former
is suppressed in the $V_{\lambda ' \lambda}$ amplitudes, Eq.
(\ref{vimpulsefinal}) by kinematical factors. However, one might think that
there could be physical situations which could be sensitive to this GPD: in
the amplitudes where $\lambda ' \neq  \lambda $ the transition with
$H^{IS}$ is done at the cost of using the D-wave of the deuteron, i.e., by
making use of angular momentum. The term with $E^{IS}$ could flip the
helicity at the nucleon level and therefore, the (rather small) D-wave
admixture in the deuteron is not necessary.

Unfortunately, most of the observables are dominated by amplitudes with
$\lambda ' =  \lambda $. We have checked that effects due to $E^{IS}$ are
negligible and we have only shown how tiny they are for the sum rules, just
for illustrative purposes. Furthermore, when modelling $E^q$ following the
steps explained in \cite{GOEKE01}, one realizes that the isoscalar
combination is suppressed. Recall that $E^q$ is normalized to the Pauli form
factor, that in the forward limit gives just the anomalous magnetic moment,
very small for the isoscalar case.

Let us stress that the available models of GPDs are fraught with
uncertainties, in particular in the ERBL region. There, GPDs describe
the emission of a $q\bar{q}$ pair from the target, and an ansatz only
using the information from usual parton densities should be used with
care.    Notice also
that, while for $x>\xi$ GPDs are bounded from above
\cite{Pire:1999nw}, no analogous constraints are known in the ERBL
region.
A particular type of contribution in the ERBL region is the
Polyakov-Weiss $D$-term \cite{Polyakov:1999gs}, which we will 
not include in our analysis.

\subsection{Results}

With the ingredients mentioned before we plot in figures \ref{hifigure} and
\ref{hitildefigure} the corresponding generalized quark distributions, which is
flavorblind for the deuteron case. The support of these functions is
$-1<x<-1$ but we have plotted only the central region. In addition, due
to the assumption made when modelling $\tilde{H}$ for the nucleon 
(the non-contribution of the polarized sea) we have that 
$\tilde{H}_i(x\leq \xi,\xi,t)$ vanishes. 

The rapid falloff of the GPDs with $x$ reflect the fact that the
impulse approximation, i.e, the single nucleon contribution cannot
account for very large longitudinal momentum. 

Notice also the huge differences in the scales of the various GPDs~: for
the vector sector, $H_3$ dominates over the others, whereas 
$H_4$ or $H_5$ are very small. The respective sizes may be related to the 
values of the different deuteron form factors for the GPDs that have a sum-rule
connection to them (see Eq.\ref{sumr}). 

The form factors that we have used in the current $(G_1,G_2,G_3)$ are
related with the usual charge monopole, $G_C$, magnetic dipole, $G_M$
and charge quadrupole, $G_Q$ in the following way~:

\begin{eqnarray}
G_1(t) & = & G_C(t)-\frac{2}{3}\eta G_Q(t) \nonumber \\
G_2(t) & = & G_M(t) \\
(1+\eta) G_3(t) & = & G_M(t) - G_C(t)+\left( 1+ \frac{2}{3}\eta
\right) G_Q(t) \nonumber
\end{eqnarray}

\noindent whith $\eta=\frac{-t}{4M^2}$. With the flavour decomposition
of the form factors for the deuteron, we have~: $G_i^u=G_i^d \equiv G_i^q
 = 3 G_i$. The dominant form factor is $G_3^q$ due mainly to the size
 of $G_Q(t)$ (see \cite{GARCON01}), and if we consider the form factors as
 a  normalization condition for the GPDs, it is natural that $H_3$
 dominates over the other GPDs. 

Notice, however, that the fact that a GPD is large does not mean
necessarily that it plays a major role in the observables~: it has to
be multiplied by the corresponding kinematical coefficients.     

It is worthwhile mentioning that we have plotted the GPDs at a
particular value of $-t$, i.e. we cannot set $t=0$ to study the $\xi$
and $x$ behaviour. The reason is that, even if we assume a factorized
form for the $t$ dependence in the nucleon, in the deuteron we cannot
isolate this $t$ dependence. In fact, there are two sources of $t$
dependence in $H_i$ and $\tilde{H}_i$~: first the explicit $t$
dependence in the nucleon GPDs and, the most important one, the
transverse momentum in the deuteron wave function. Then, we cannot
circumvent the kinematical relationship between $\xi$ and $t$, 
Eq. (\ref{xitrelation}) and for a non-vanishing $\xi$ we have
inevitably a non-vanishing $t$.

\begin{figure}   
\resizebox{0.50\textwidth}{!}{%
\includegraphics{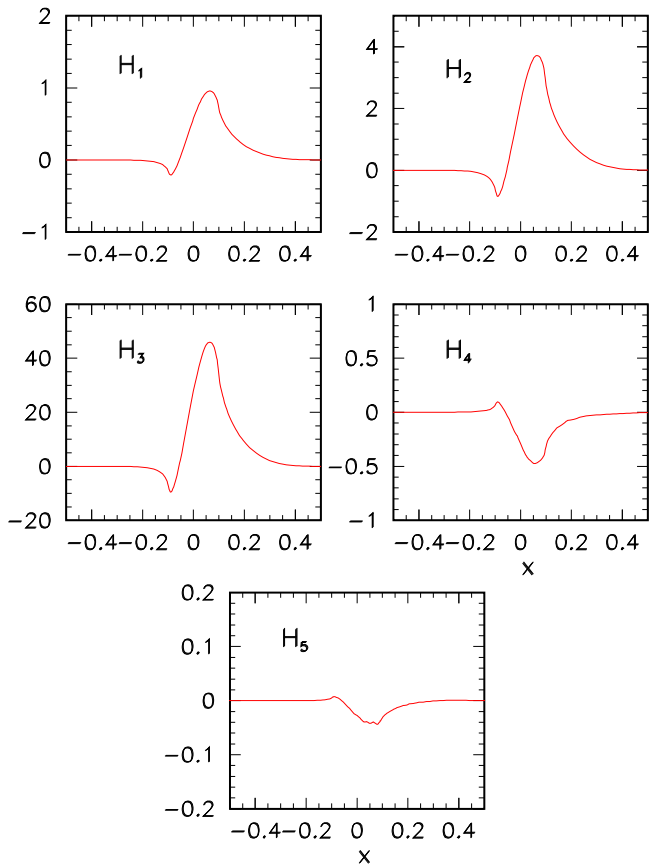}
}
\caption{ Generalized Quark Distributions for the deuteron at $Q^2= 2$
  GeV$^2$, $\xi=0.1$  and $t=-0.25$ GeV$^2$.}
\label{hifigure}
\end{figure}

\begin{figure} 
\resizebox{0.50\textwidth}{!}{%
\includegraphics{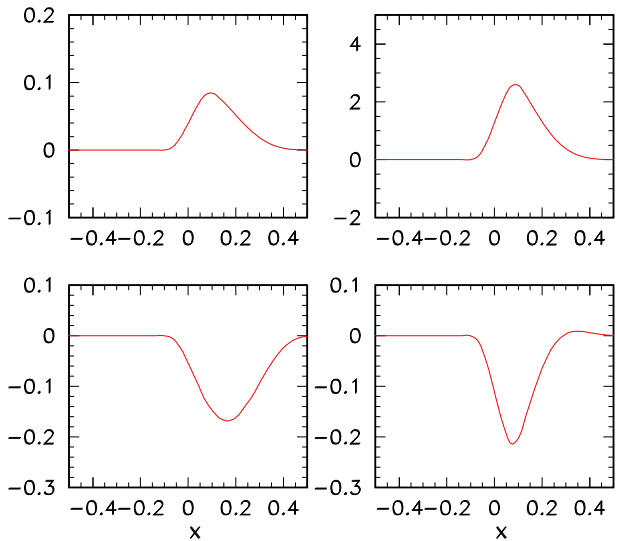}
}
\vskip -4cm
\caption{ Generalized polarized Quark Distributions for the deuteron
  at $Q^2= 2$ GeV$^2$, $\xi=0.1$  and $t=-0.25$ GeV$^2$. $\tilde{H}_1$
  (upper-left),  $\tilde{H}_2$ (upper-right), $\tilde{H}_3$
  (lower-left), $\tilde{H}_4$ (lower-right), }
\label{hitildefigure}  
\end{figure}

\subsection{Sum Rules~: tests and discussions}

In the impulse approximation we have retained only the lowest
Fock-space state of the deuteron (Fig. \ref{impulsescheme}, a and
b). As we see, the $q\bar{q}$ components which are tested in the
region $|x| < \xi$, are considered only within the nucleon itself
(Fig. \ref{impulsescheme}.b). 

We have also neglected the $N\bar{N}$ components in the deuteron wave function,
which could give rise to diagrams like the one in
Fig. \ref{impulsescheme}.c.
When one evaluates the elastic deuteron form factor
\cite{FRANKFURT79,FRANKFURT93} this is an exact approximation since
one can always choose a frame where the momentum transfer is purely
transverse and in that case, $\Delta^+=0$ and no pairs can be created
from or annihilated into a photon.

In the DVCS and DEMP cases there is a non-vanishing momentum transfer in
the longitudinal direction, controled by the skewness parameter
$\xi$. Therefore, there are necessarily diagrams where the final photon goes
out from the anhilation of, for example, a $N\bar{N}$ pair or a
$q\bar{q}$(see figure  \ref{impulsescheme}.c).
One has to include these higher Fock-space components to recover Lorentz
invariance.

 Lorentz invariance is the physical reason why the sum rules (\ref{sumr}),
obtained by integrating the GPDs over $x$, become $\xi$-independent. When we
perform that integration we have matrix elements of local operators  (form factors) that cannot
depend on the artifacts of the kinematics, i.e. of the separation between
transverse and longitudinal momentum transfer (see \cite{BRODSKY01} for a
more detailed discussion on this point).

 We can make use of this relationship between the $\xi$-independence of the sum
rules and the contribution of higher Fock-space states in the deuteron to
check how accurate the impulse approximation is. We have plotted in figure
\ref{sumrulesxidep} the quantities:

\begin{equation}
I_i(\xi)=\int_{-1}^1 dx H^q_i(x,\xi,t)
\end{equation}

For a fixed $t$, the functions $I_i(\xi)$ would be  constant if the
impulse approximation was  exact : straight lines in the
figure show the 'theoretical' values of the sum rule according to the
experimental parameterization of the form factors. Any residual $\xi$
dependence of $I_i(\xi)$ is a measure of the importance of the higher
Fock-space states that we have not included in our description. Looking at
this figure and to the corresponding one for the axial case, we can see
that this dependence is fairly mild, which indicates that, in the
kinematical regime that we are interested in, the deuteron is essentially a
two-nucleon state\footnote{We checked that the
inclusion of the nucleon GPD $E^{IS}$ does not introduce any improvement at
all:  its contribution vanishes exactly at $\xi=0$, and  is
always small at other values of $\xi$¥.}¥. As $\xi$ increases the impulse approximation 
would become a too rough approximation with respect to Lorentz invariance. 

One should distinguish between the variation in $\xi$ of the quantities
$I_i(\xi)$ and the particular values they take, which are sensitive also to
the details of the employed model. In figure  \ref{sumrulesxidep} we see
that the points obtained with our calculations are quite close to the
experimental parameterization. Obviously, the models works better at smaller
$\xi$, for the reasons exposed above. In fact, in figure \ref{tdepSR}, we show
the t-dependence of the sum rules at $\xi=0$, where it is clearly seen that
results agrees quite well with the experimental parameterization, when
available, or with the values imposed by time reversal or Lorentz
invariance. This comes as no surprise since it is well known that the
light-cone deuteron wave function is able to give the deuteron form factors
at the momentum transfer we are working (see \cite{COOKE} for a recent
review). In the context of our discussion, at $\xi=0$ the pair creation or
annihilation with the photon vanishes in the light-cone formalism.

One final remark concerns also the subtleties of the light-cone formalism:
nucleons are on-shell, i.e., they verify, with the notation employed in the
previous section, that:

\begin{equation}
p_i^- = \frac{m^2 + \mathbf{p_{i_\perp}^2}}{2 p_i^+}
\end{equation}

But, they are off light-cone energy shell, and as a consequence, if
$P^+=p_1^+ + p_2^+$ and $\mathbf{P_\perp} =  \mathbf{p_{1_\perp}} +
\mathbf{p_{2_\perp}}$ (as it is the case), one has that $P^- \neq p_1^- +
p_2^-$. Therefore, one has that the momentum transfer  $t$ defined from the deuteron
variables does not coincide with the one defined from the variables of the
active nucleon. Moreover, the upper limit over $\xi_N^2$ is not $\frac{-t}{4
m^2 -t}$. If nevertheless one enforces $\xi_N^2$ to have this upper limit,
this leads only to a tiny shift in the value of $\alpha_{min}$ in the
integrals. From the practical point of view, these differences are
too tiny to be seen in the numerical calculation, unless one goes to very
large values of $-t$. This just reflects the fact that the off-shellness
effects in the light-cone energy are of the order of the binding energy over
the longitudinal momentum \cite{BRODSKY89}:

\begin{equation}
P^- - (p_1^- + p_2^-) \propto \frac{V}{P^+}
\end{equation}

\noindent and in our case this is of the order of the binding energy of the
deuteron over the center of mass energy, i.e, very small.

\begin{figure}
\resizebox{0.50\textwidth}{!}{%
\includegraphics{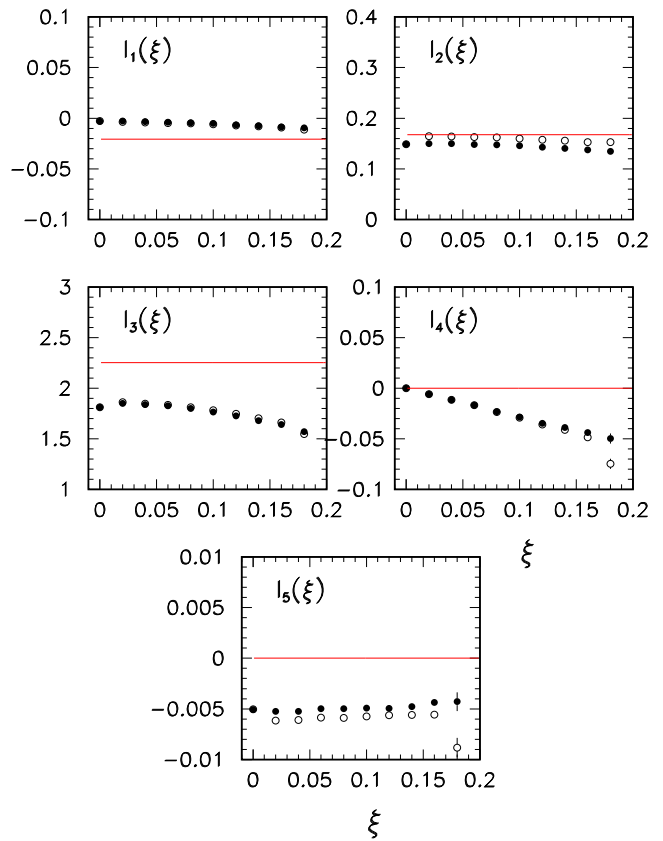}
}
\caption{ Sum rules for the vector GPDs at $Q^2= 2$ GeV$^2$ and
  $t=-0.5$ GeV$^2$. Solid lines are the expected theoretical result
  and points are the results obtained with our model. Filled points~:
  the nucleon GPD $E^{\mbox {\scriptsize IS}}$ is not included~; Empty
  points~: $E^{\mbox {\scriptsize IS}}$ included. }
\label{sumrulesxidep}
\end{figure}

\begin{figure}
\resizebox{0.50\textwidth}{!}{%
\includegraphics{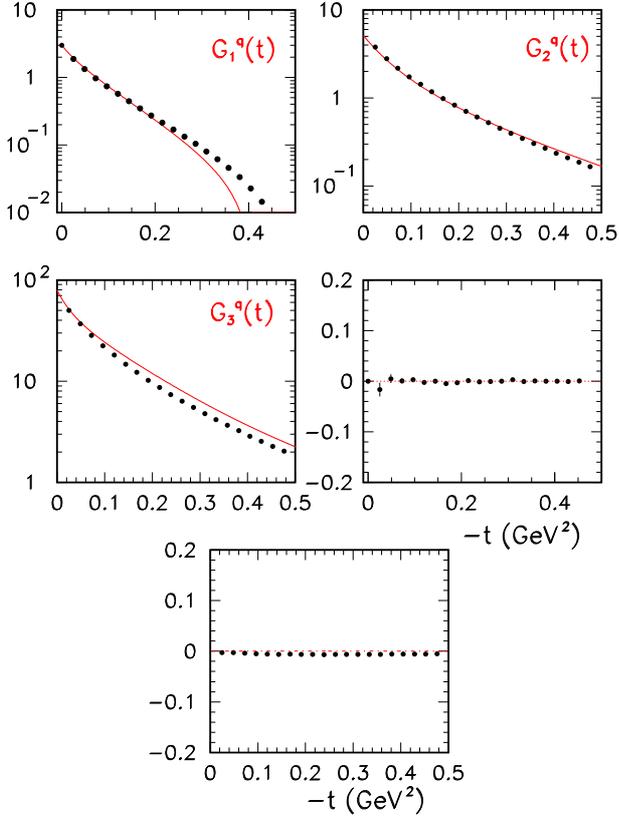}
}
\caption{ Sum rules for the vector GPDs at $Q^2= 2$ GeV$^2$ and fixed
  $\xi=0$ as a function of t. Solid lines represent the expected
  theoretical values whereas points are the results of our evaluation.}
\label{tdepSR}¥
\end{figure}

\begin{figure}
\resizebox{0.50\textwidth}{!}{%
\includegraphics{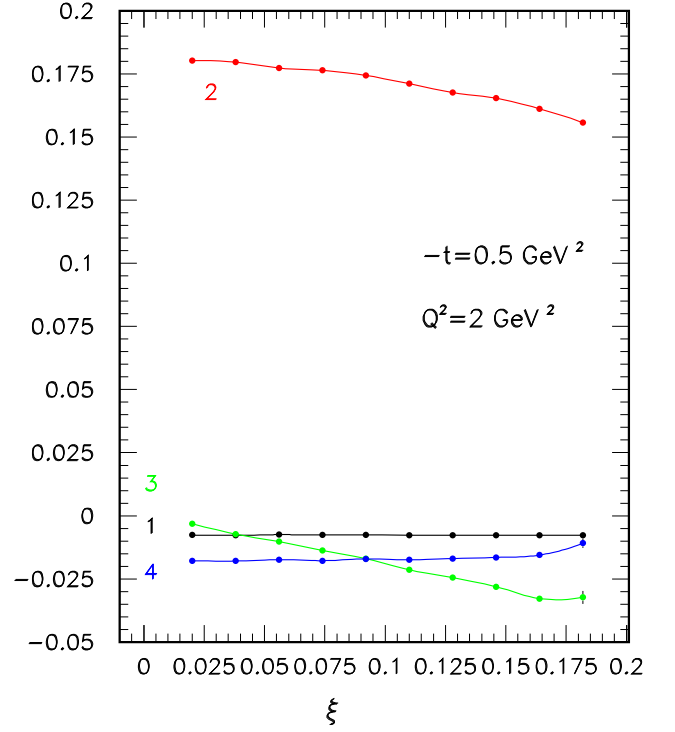}
}
\caption{ Sum rules for the axial-vector GPDs at $Q^2= 2$ GeV$^2$ and
 $t=-0.5$ GeV$^2$.}
\end{figure}

\section{DVCS Amplitudes and Cross Sections}
There are two processes that contribute to the deeply virtual Compton scattering amplitude of 
Eq. 1 under consideration.
The first one is the Bethe-Heitler process where the outgoing photon is produced
from the lepton line. Its amplitude (for either electrons or positrons) is given by
\begin{equation}
T_{\mbox{\scriptsize BH}} = - \frac{e^3}{t} \;
\epsilon^*_\mu(\vec{q}\,',\lambda') j_\nu(0) L^{\mu \nu}
\end{equation}

\noindent where $e$ is the proton charge.
The deuteron current is given by:

\begin{eqnarray}
j_{\mu} = &-& G_1(t) ( \epsilon'^* \cdot \epsilon) 2 \bar{P}_\mu + G_2(t)
\left[ (\epsilon'^* \cdot 2 \bar{P}) \epsilon_\mu + (\epsilon \cdot 2 \bar{P})
\epsilon'^*_\mu \right] \nonumber \\
&-& G_3(t) (\epsilon'^* \cdot 2 \bar{P})(\epsilon \cdot 2 \bar{P})
\frac{ \bar{P}_\mu}{ M^2}
\end{eqnarray}
\noindent where the three form factors have been
measured in the low and medium momentum transfer ranges\cite{GARCON01}. The 
leptonic tensor $L^{\mu \nu}$ is given by

\begin{equation}
L^{\mu \nu} = \bar{u}(\vec{k}\,',h') \left[ \gamma^\mu \frac{1}{(\not{k} '
+ \not{q} ')} \gamma^\nu  +  \gamma^\nu \frac{1}{(\not{k}- \not{q} ')}
\gamma^\mu\right] u(\vec{k},h)
\end{equation}
The Bethe-Heitler process is thus completely known in terms of already
measured form factors. 

The second process where the photon is emitted from the hadronic part is
more interesting in terms of the study of the hadronic structure. It is
called virtual Compton scattering since it can be decomposed in a
$\gamma^* A \to \gamma B $ process. Its amplitude $T_{\mbox{\scriptsize
VCS}}$ is written as
\begin{equation}
T_{\mbox{\scriptsize VCS}} =
\pm
\frac{e^3}{Q^2}
\;
\sum_{\lambda} \Omega(h,\lambda) M_{H'\lambda',H\lambda}
\end{equation}

\noindent
where the upper sign is for electrons and the lower one for positrons and
the function $\Omega$ comes from the decomposition of the leptonic current
in terms of the polarization vector of the virtual photon

\begin{eqnarray}
&&\bar{u}(\vec{k}\,',h)\gamma^\nu u(\vec{k},h)  =  \sum_{\lambda}
\frac{Q}{\sqrt{1-\epsilon}}\tilde{\Omega}(h,\lambda) \epsilon^\nu(\vec{q},\lambda) \\
&&\tilde{\Omega}(h,\lambda)  =   \left[ \delta_{\lambda
0} \sqrt{2 \epsilon} - \frac{\lambda}{\sqrt{2}} (\sqrt{1+ \epsilon} + 2 h
\lambda \sqrt{1- \epsilon}) e^{-i \lambda \phi} \right] \nonumber
\end{eqnarray}

Sometimes we will also use:

\begin{equation}
\Omega(h,\lambda) \equiv \frac{Q}{\sqrt{1-\epsilon}} \tilde{\Omega}(h,\lambda)
\end{equation}

The photon-deuteron helicity amplitudes are defined as:

\begin{equation}
M_{H'\lambda ',H\lambda} = \epsilon^*_\mu(\vec{q}\,',\lambda')
\epsilon_\nu(\vec{q}\, ,\lambda) H^{\mu \nu}
\end{equation}

\noindent and
\begin{eqnarray}
H^{\mu \nu} & = &  ( g^{\mu \nu} - \tilde{p}^\mu \tilde{n}^\nu -
\tilde{n}^\mu \tilde{p}^\nu)   \nonumber \\
& & \int_{-1}^{1} dx \left(
\frac{1}{x-\xi+i\eta} +\frac{1}{x+\xi-i\eta}\right) \sum_q e_q^2
V^q_{\lambda ' \lambda} \nonumber \\
& + & i \epsilon^{\mu \nu \alpha \beta} \tilde{p}_\alpha
\tilde{n}_\beta  \\
& &\int_{-1}^{1} dx \left(
\frac{1}{x-\xi+i\eta}-\frac{1}{x+\xi-i\eta}\right) 
\sum_q e_q^2
A^q_{\lambda ' \lambda}  \nonumber
\end{eqnarray}

\noindent with the convention $\epsilon_{0123}=+1$.

For completeness, let us first write down the  formula for the cross
section of the Bethe Heitler process on the deuteron
\begin{eqnarray}
\overline{\sum} |T_{\mbox{\scriptsize BH}}|^2 & = & \frac{(4 \pi
\alpha_{em})^3}{t^2} [{\cal K}_A A(t) + {\cal K}_B B(t) ]
\end{eqnarray}

\noindent where $A$, $B$ are the elastic structure functions of the
deuteron, which are well known in the low momentum region, and the kinematical 
coefficients are:

\begin{eqnarray}
{\cal K}_B & = & -\frac{2 M^2}{(k \cdot q'\,)(k' \cdot q'\,)} [(2 (k \cdot
q'\,) + t)^2 + (2 (k \cdot q'\,) + Q^2)^2] \nonumber\\
{\cal K}_A & = & - {\cal K}_B + \frac{4 t}{(k' \cdot q'\,)} (M^2 + s + Q^2
- 2 s_{kp}) \nonumber \\
& + & \frac{t}{2 (k \cdot q'\,)(k' \cdot q'\,)} \left\{(2 M^2 + Q^2 - 2
s_{kp})^2 - (Q^2 - 2 t)^2 \right. \nonumber \\ &   &  \left. + 4 (Q^2 + s -
s_{kp})^2 + 4 t(t + s - s_{kp}) \right\}
\end{eqnarray}

\noindent
with $s_{kp}=(k +p)^2$.

The VCS amplitude gives a contribution to the cross section which may be
decomposed in terms of its azimuthal dependence as
\begin{eqnarray}
& &\overline{\sum}|T_{\mbox{\scriptsize VCS}}|^2  = 
\frac{1}{3}
\frac{(4 \pi \alpha_{\mbox {\scriptsize em}})^3}{Q^2 (1-\epsilon)}
\sum_{H,H'} \left(2 |M_{H'1,H 1}|^2 + 2 |M_{H'1,H -1}|^2 \nonumber \right. \\
 & + &
4 \epsilon |M_{H'1,H 0}|^2 \nonumber\\
& + & 4 \sqrt{\epsilon (1+\epsilon)} \; \cos \phi \;
{\mbox{Re}}[ M_{H' 1, H 0} M^*_{H' 1,H -1} -M_{H' 1, H 0} M^*_{H' 1,H
1} ] \nonumber \\
& - & \left.
4 \epsilon \cos (2\phi) \; {\mbox{Re}}[M_{H' 1, H -1} M^*_{H' 1,H 1}]\right).
\end{eqnarray}

The interference between the two processes leads to a contribution to the DVCS cross
section which may be written as
\begin{eqnarray}
\overline{\sum}
&  &\left(T_{\mbox{\scriptsize VCS}} T_{\mbox{\scriptsize BH}}^*
+ T_{\mbox{\scriptsize VCS}}^* T_{\mbox{\scriptsize BH}}\right)   =
\mp \frac{2}{3}
\frac{(4 \pi \alpha_{\mbox {\scriptsize em}})^3}{Q t \sqrt{1-\epsilon}}
\nonumber \\
 & \cdot & \sum_{H,H',\lambda,h} 2 h \;{\mbox
{Re}}[\epsilon^*_\mu(\vec{q}\,',\lambda'=+1) j_\nu
L^{\mu\nu}\tilde{\Omega}(h,\lambda)] \nonumber \\ & \cdot &
{\mbox {Re}}[M_{H'1,H \lambda}]
\end{eqnarray}

\noindent where the upper sign stands for electrons and the lower one for
positrons.

The study of the initial electron helicity dependence may be expressed
through  the following weighted contributions to the cross section
\begin{eqnarray}
\overline{\sum}2 h |&  &T_{\mbox{\scriptsize BH}}|^2 = 0
\end{eqnarray}

\begin{eqnarray}
&  &\overline{\sum}2 h  |T_{\mbox{\scriptsize VCS}}|^2 =
\frac{4}{3}
\frac{(4 \pi \alpha_{\mbox {\scriptsize em}})^3}{Q^2}
\sqrt{\frac{\epsilon}{1-\epsilon}}  \\ & &
\sum_{H,H'} \; \sin \phi \;
{\mbox{Im}}[ M_{H' 1, H 0} M^*_{H' 1,H 1} -M_{H' 1, H 0} M^*_{H' 1,H
-1} ] \nonumber
\end{eqnarray}

\begin{eqnarray}
&  &\overline{\sum} 2 h
\left(T_{\mbox{\scriptsize VCS}} T_{\mbox{\scriptsize BH}}^* +
 T_{\mbox{\scriptsize VCS}}^* T_{\mbox{\scriptsize BH}}\right)   =
\mp \frac{2}{3}
\frac{(4 \pi \alpha_{\mbox {\scriptsize em}})^3}{Q t \sqrt{1-\epsilon}}
\nonumber \\
 & \cdot & \sum_{H,H',\lambda,h} 2 h \;{\mbox
{Im}}[\epsilon^*_\mu(\vec{q}\,',\lambda'=+1) j_\nu  L^{\mu\nu}
\tilde{\Omega}(h,\lambda)] \nonumber \\ & \cdot &
{\mbox {Im}}[M_{H'1,H \lambda}]
\end{eqnarray}

\noindent where the upper sign stands for electrons and the lower one for
positrons.

\section{Numerical Results for DVCS}

Our model enables us now to estimate the cross section of coherent deeply virtual Compton
 scattering on the deuteron. We are
particularly interested by the forthcoming experiments at JLab and Hermes
at DESY, and we thus shall present results for the kinematics of these experimental
set ups.

\begin{figure}[t]
\resizebox{0.50\textwidth}{!}{%
\begin{tabular}{lr}
\includegraphics{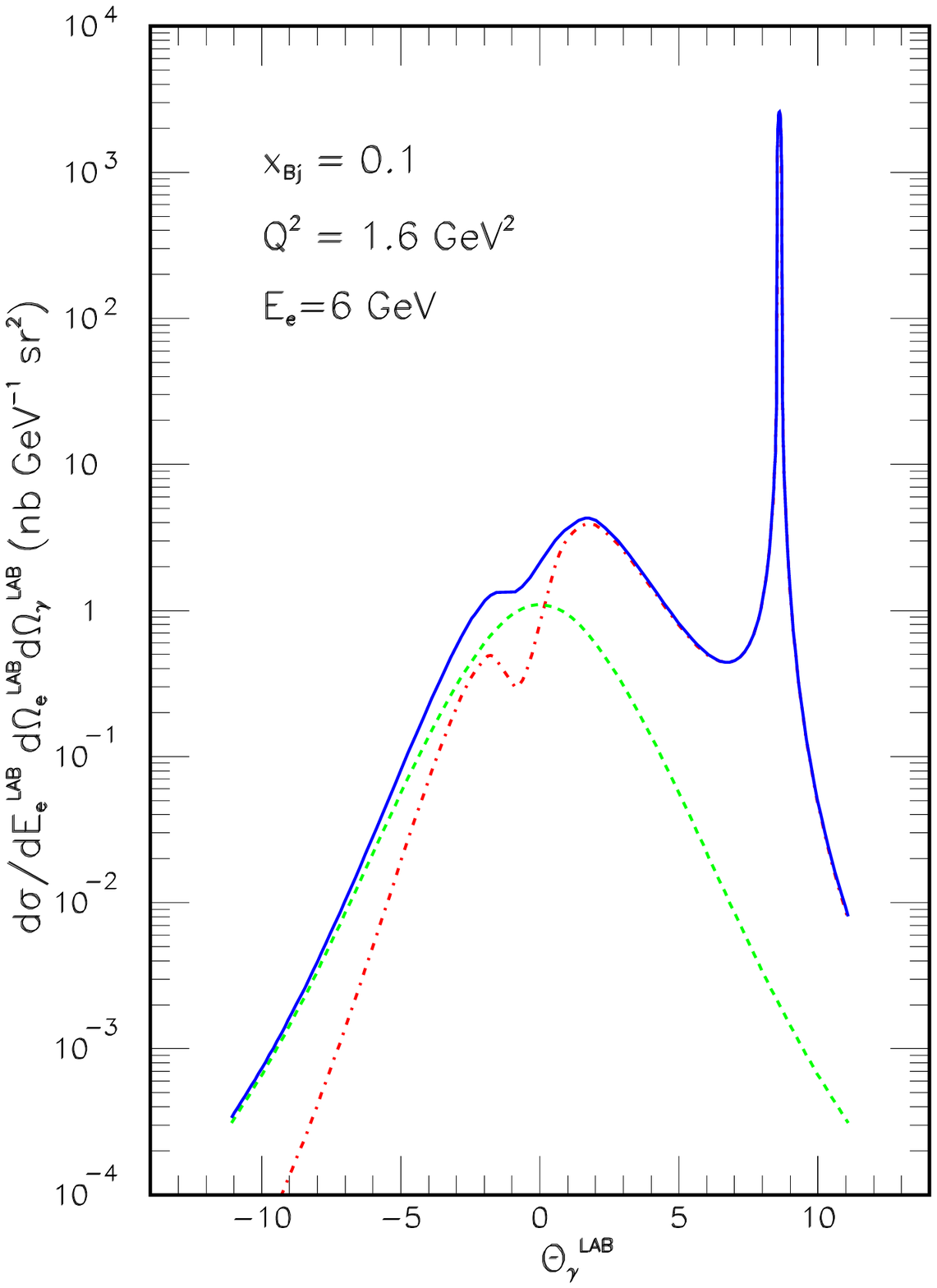} &
\includegraphics{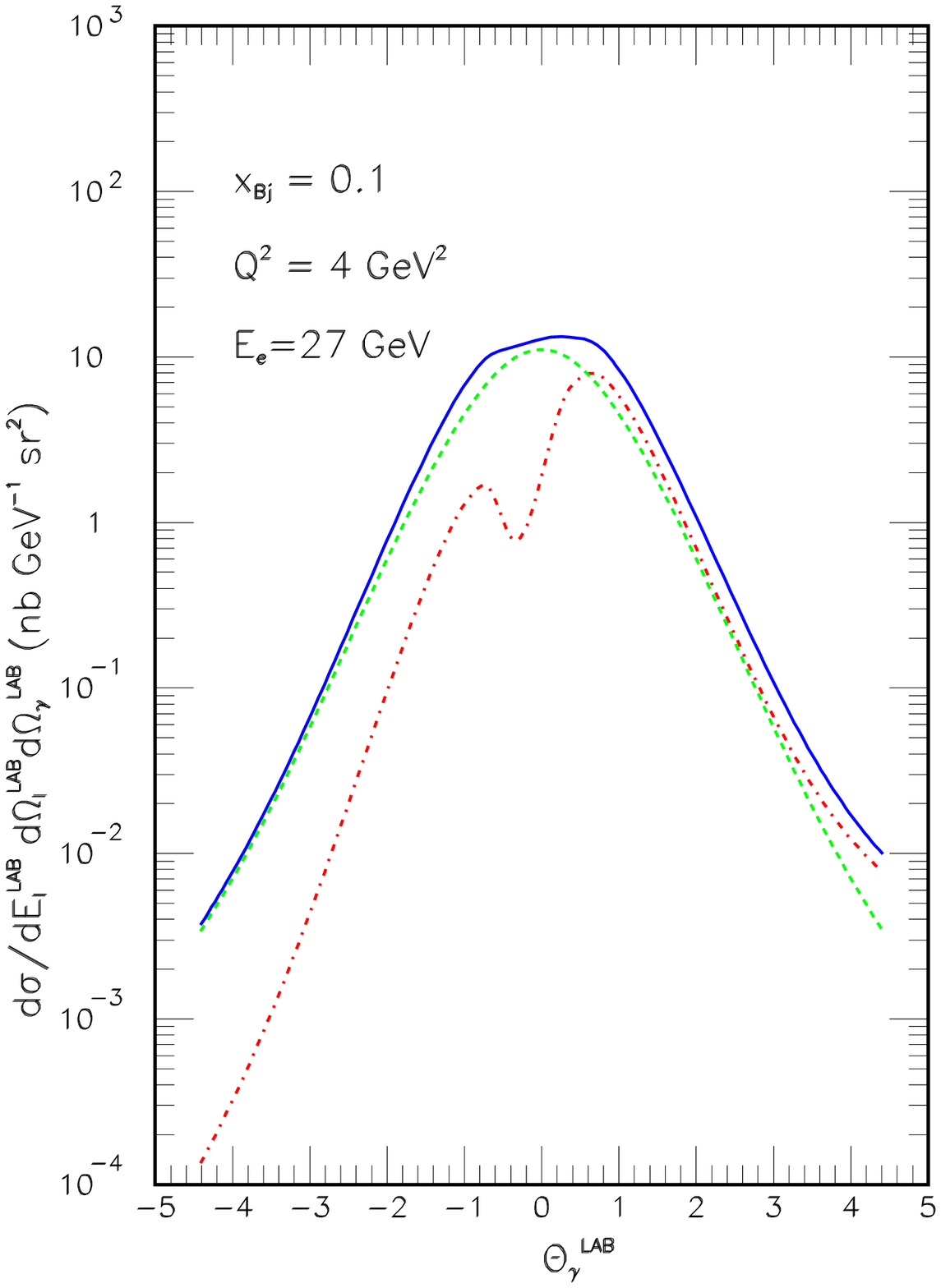}
\end{tabular}
}
\vskip -0.5cm
\caption{Unpolarized Differential Cross section for DVCS for typical
kinematics at JLab (left panel) and HERMES (right panel).
Dashed-dotted line: BH only; dashed line: DVCS only; full
line: BH + DVCS + Interference.}
\label{csec}
\end{figure}

We present in Fig. \ref{csec} the unpolarized cross section for low (left panel) and medium 
(right panel) energy reactions. The Bethe-Heitler and VCS contributions are shown as well as 
their  interference. The relative importance of these contributions depend much on the 
production angle of the final photon, as can be read from the figure. To discuss the feasibility
of the experiment, a comparison to the proton target case is welcome. This is shown on 
Fig \ref{pvsd} for medium energy reactions.

\begin{figure}
\resizebox{0.50\textwidth}{!}{%
\includegraphics{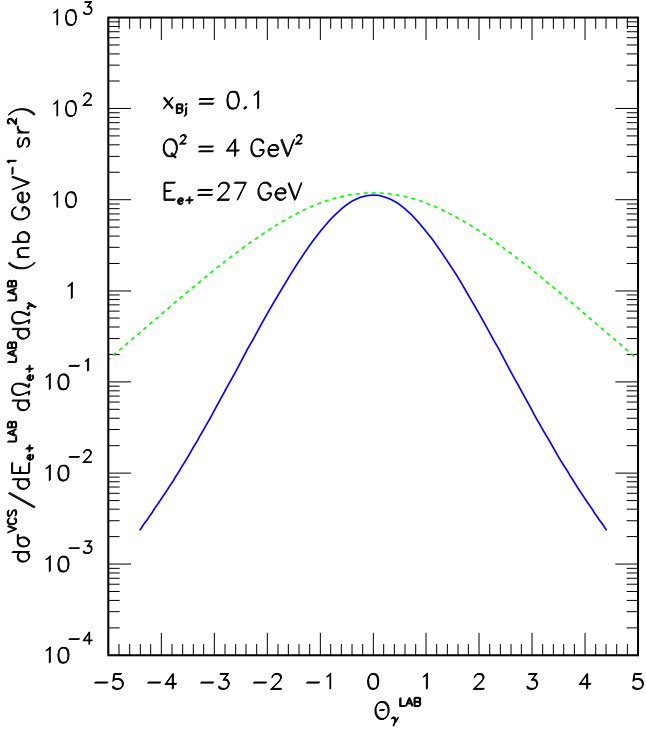}
}
\caption{Comparison of the DVCS cross sections on a proton (dashed
  line) and deuteron (solid line) target for 
 $\xbj=0.1$ in the deuteron case (resp. 0.2 for the proton), $Q^2= 4$ GeV$^2$ and $E_{e^+}=
27$ GeV.}
\label{pvsd}
\end{figure}

Coherent deep VCS is certainly not a negligible effect at small values of $t$ and we can 
expect that this process should  soon 
become observable so that some knowledge of the deuteron GPDs will  become accessible.
More than testing the validity of the impulse approximation, the goal of such an experiment is 
to observe some definite deviation from the impulse approximation predictions,
 thereby indicating some non-trivial short distance content
of the deuteron. To scrutinize such effects, it is interesting to turn to some more specific
observables, such as spin and charge asymmetries.

  The beam spin asymmetry is defined as

\begin{equation}
A_{LU} (\phi) = \frac{d\sigma^\uparrow (\phi) - d\sigma^\downarrow
(\phi)}{d\sigma^\uparrow (\phi)+ d\sigma^\downarrow
(\phi)}
\end{equation}
where $\phi$ is the angle between the lepton and hadron scattering planes.
The numerator is proportional to the
interference between the Bethe-Heitler and the VCS amplitudes.

Our predictions calculated with our modelized
deuteron GPD's are shown on Fig. \ref{asy} for JLab and Hermes energies. The sign
of the asymmetry is reversed for a positron beam. Such a sizable asymetry should be quite 
easily measured. It will constitute a crucial test of the validity of any model.

It has been shown\cite{DGPR}¥ that, asymptotically, the beam spin asymmetry exhibits a
$\sin(\phi)$ azimuthal dependence. We have performed a Fourier
decomposition
of the asymmetry $A_{LU}$ obtained for the deuteron and, indeed, we have checked the
dominance of the $\sin(\phi)$ component, even at relatively low values of
$Q^2$. But we also have a sizeable  $\sin(2 \phi)$ component (see
figure (\ref{ssaf})), which  is less suppressed than in the nucleon case. This is likely to 
come from the following fact :
the Bethe-Heitler propagators  
 exibit  when $t \neq t_0$ an azimuthal dependence  in $cos(\phi)$ which goes with terms of the
order of $t/Q^2$, but also with terms of the order $m^2/Q^2$. In the nucleon case this
 $sin(2\phi)=2 sin(\phi) cos(\phi)$- 
component was already seen\cite{HERMES01} under some kinematical conditions, and 
the larger deuteron mass enhances it. In that respect, 
we expect that the  scaling regime signed by the dominance of the 
$\sin(\phi)$ component\cite{DGPR} is likely to be reached later in the deuteron
case than in the proton case.

\begin{figure}
\resizebox{0.50\textwidth}{!}{%
\includegraphics{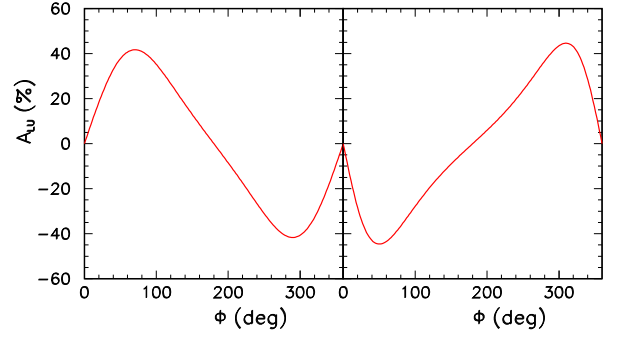}
}
\caption{Azimuthal dependence of the Beam Spin Asymmetry as defined in
the text. Left: $\xbj=0.2$, $Q^2= 2$ GeV$^2$ and $E_e=
6$ GeV. Right: $\xbj=0.1$, $Q^2= 4$ GeV$^2$ and $E_{e^+}=
27$ GeV. In both cases $t$ is fixed to $-0.3$ GeV$^2$.}
\label{asy}
\end{figure}

\begin{figure}[t]
\resizebox{0.50\textwidth}{!}{%
\begin{tabular}{lr}
\includegraphics{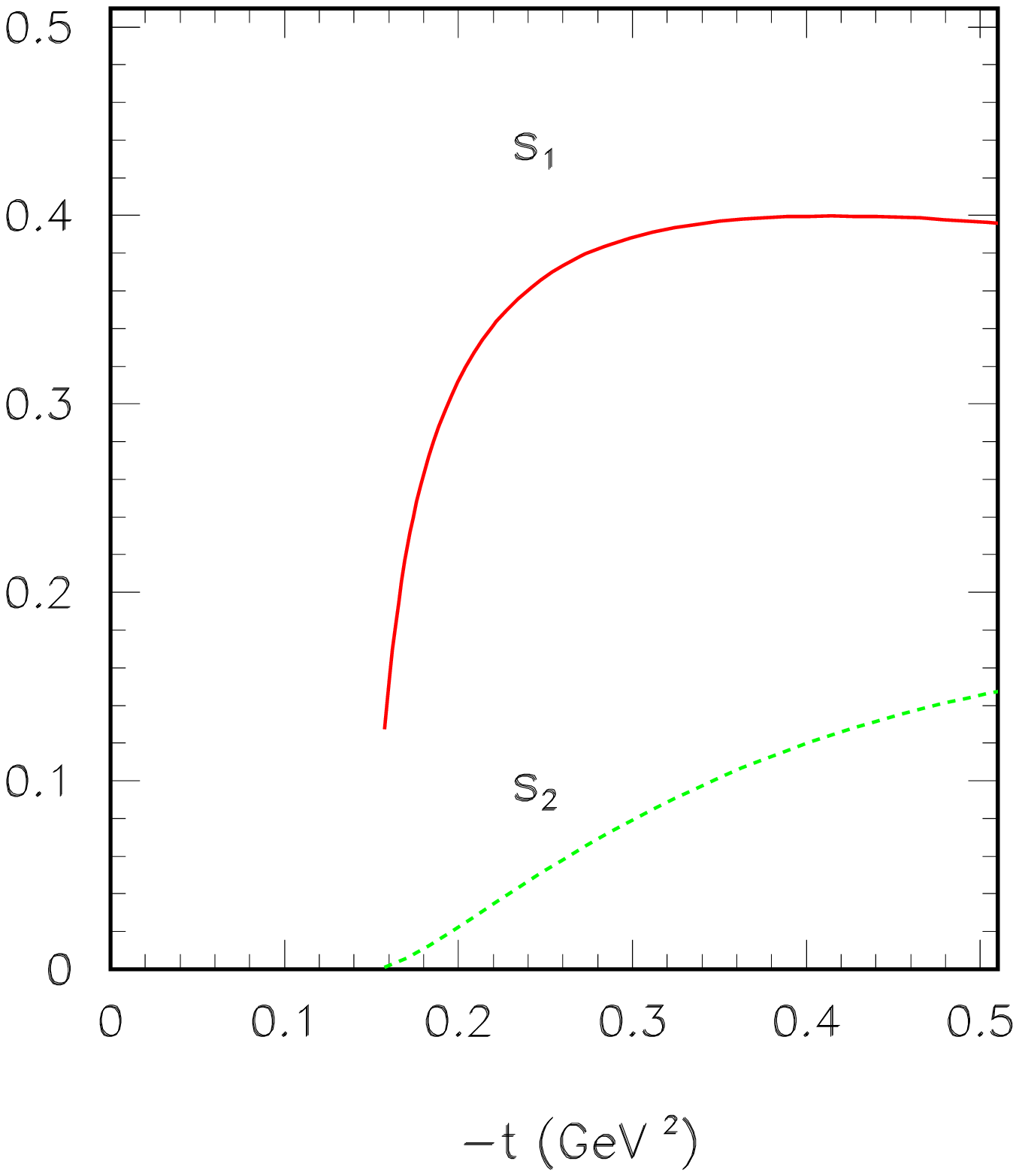} &
\includegraphics{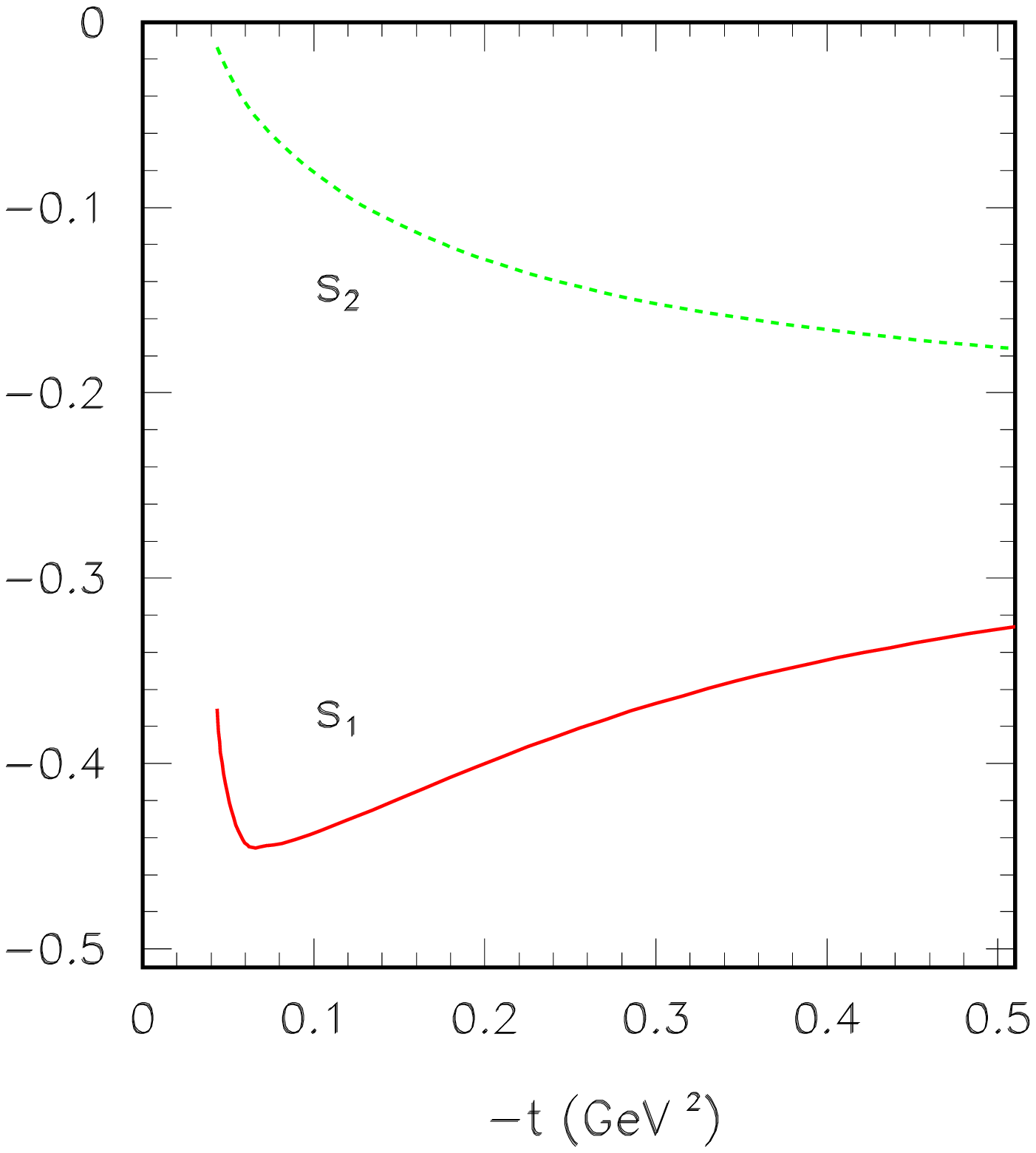}
\label{ssaf}¥
\end{tabular}
}
\caption{Coefficients of the Fourier decomposition of the beam spin
  asymmetry~: $A_{LU} = a_0 + s_1 \sin{\phi} + s_2 \sin{2 \phi}$ as a
  function of t. Values of $\xbj$, $Q^2$ and $E_l$ as  in
  Fig. \ref{asy}.}
\end{figure}

The beam charge asymmetry
\begin{equation}
A_{C} (\phi) = \frac{d\sigma^{e^+} - d\sigma^{e^-}} 
{d\sigma^{e^+} + d\sigma^{e^-}}
\end{equation}
is also proportionnal to the interference of the Bethe-Heitler and the 
VCS processes. Its characteristic azimuthal dependence is shown on Fig{\ref{CA}}.
Its size is large enough for a feasible experimental evaluation.

\begin{figure}[t]
\resizebox{0.50\textwidth}{!}{%
\includegraphics{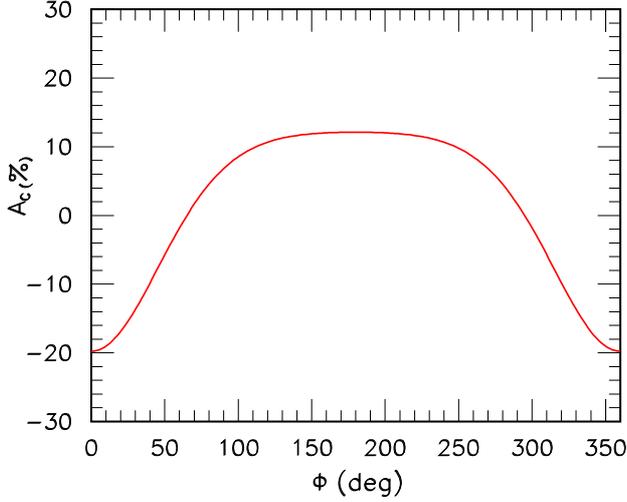} 
}
\caption{Azimuthal dependence of the beam charge asymmetry $A_C$ for
  the kinematics of HERMES shown in previous figures.}
\label{CA}
\end{figure}

\section{Deep exclusive meson electroproduction}
Deep exclusive meson electroproduction\cite{VDH99} may be discussed along the same lines as the DVCS
reaction. The same QCD factorization property exists which allows to separate a short distance 
subprocess from long-distance matrix elements, provided the initial virtual photon is longitudinally
 polarized. The same GPDs appear in principle in the
 deuteron sector, but selection rules select some of the GPDs.
The meson production is described within the well established colinear 
approximation\cite{ERBL}  through 
the introduction of a distribution amplitude $\Phi(z,Q^2)Ç$ which is generally parametrized 
through its asymptotic (in the sense of its $Q^2$Ç evolution) expression.
\begin{equation}
\Phi^{\rho}(z,Q^2)Ç = 6f^{\rho}z(1-z)
\end{equation}
for the $\rho$ meson, and
\begin{equation}
\Phi^{\pi}(z,Q^2) = 6 \sqrt{2}Çf^{\pi}z(1-z)
\end{equation}
for the $\pi$ meson, with $f^{\rho}= 216$ MeV and  $f^{\pi}= 92$ MeV.
The resulting amplitudes read ($\lambda,\lambda'$ denoting deuteron polarizations as above)Ç:
\begin{eqnarray}
{\cal M}^{\rho}_{\lambda',\lambda} &=& -ie \frac{16\pi\alpha_{S}}{9}
 \frac{1}{Q}\int_{0}^1dz \frac{\Phi_{\rho}(z)}{z}\nonumber\\
 &&\int_{0}^1dx ( \frac{1}{x-\xi+i\epsilon}+ \frac{1}{x+\xi-i\epsilon} )
\frac{1}{\sqrt{2}} V^q_{\lambda',\lambda}
\end{eqnarray}
for the vector meson case, and

\begin{eqnarray}
{\cal M}^{\pi}_{\lambda',\lambda} &=& -ie \frac{16\pi\alpha_{S}}{9} \frac{1}{Q}\int_{0}^1dz 
\frac{\Phi_{\pi}(z)}{z} \nonumber \\
&&\int_{0}^1dx ( \frac{1}{x-\xi+i\epsilon} + \frac {1}{x+\xi-i\epsilon}) 
\frac{1}{\sqrt{2}} A^q_{\lambda',\lambda}
\end{eqnarray}
for the pseudoscalar meson case, where the isosinglet nature of the deuteron has been used to 
simplify the results. The factor $\frac{1}{\sqrt{2}}$ in front of the
matrix elements $V_{\lambda ' \lambda}$ and $A_{\lambda ' \lambda}$
come from the flavour decomposition of the $\rho^0$ and the  $\pi^0$~,
i.e., $\frac{1}{\sqrt{2}}(|u\bar{u} \rangle - |d\bar{d} \rangle)$. 

An interesting feature of meson electroproduction is the absence of a competing subrocess such
 as the Bethe-Heitler process in DVCS. Rates are also higher than in the DVCS case by a factor
  of  $\alpha_{S}/ \alpha_{em}$. The effective
strong coupling constant has been taken as $\alpha_{S} = 0.56$ as advocated 
in \cite{VDH99}. 
 Vector meson (mostly $\rho^{0}$) production selects the vector GPDs $H^{i}$ while
  pseudoscalar meson (mostly $\pi^{0}$) production selects the axial ones $\tilde H^{i}Ç$.
  Meson pair production, described in the formalism of the generalized distribution 
  amplitudes\cite{GDA}, may also be calculated in the same way.

  We show on Fig. \ref{rhopi} the prediction of our model for    $\rho^{0}$ and  $\pi^{0}$ 
   electroproduction for
$Q^2 = 4$ GeV$^2$, $\xbj = 0.1$ and for
$Q^2 = 3$ GeV$^2$, $\xbj = 0.2$.
As in the proton case, the vector meson production is enhanced with respect to the pseudoscalar 
meson production. The pseudoscalar production is quite small since the isosinglet nature of the 
deuteron forbids any enhanced $\tilde E$ contribution due to $\pi^0$ exchange in the t-channel.

\begin{figure}[t]
\resizebox{0.50\textwidth}{!}{%
\begin{tabular}{lr}
\includegraphics{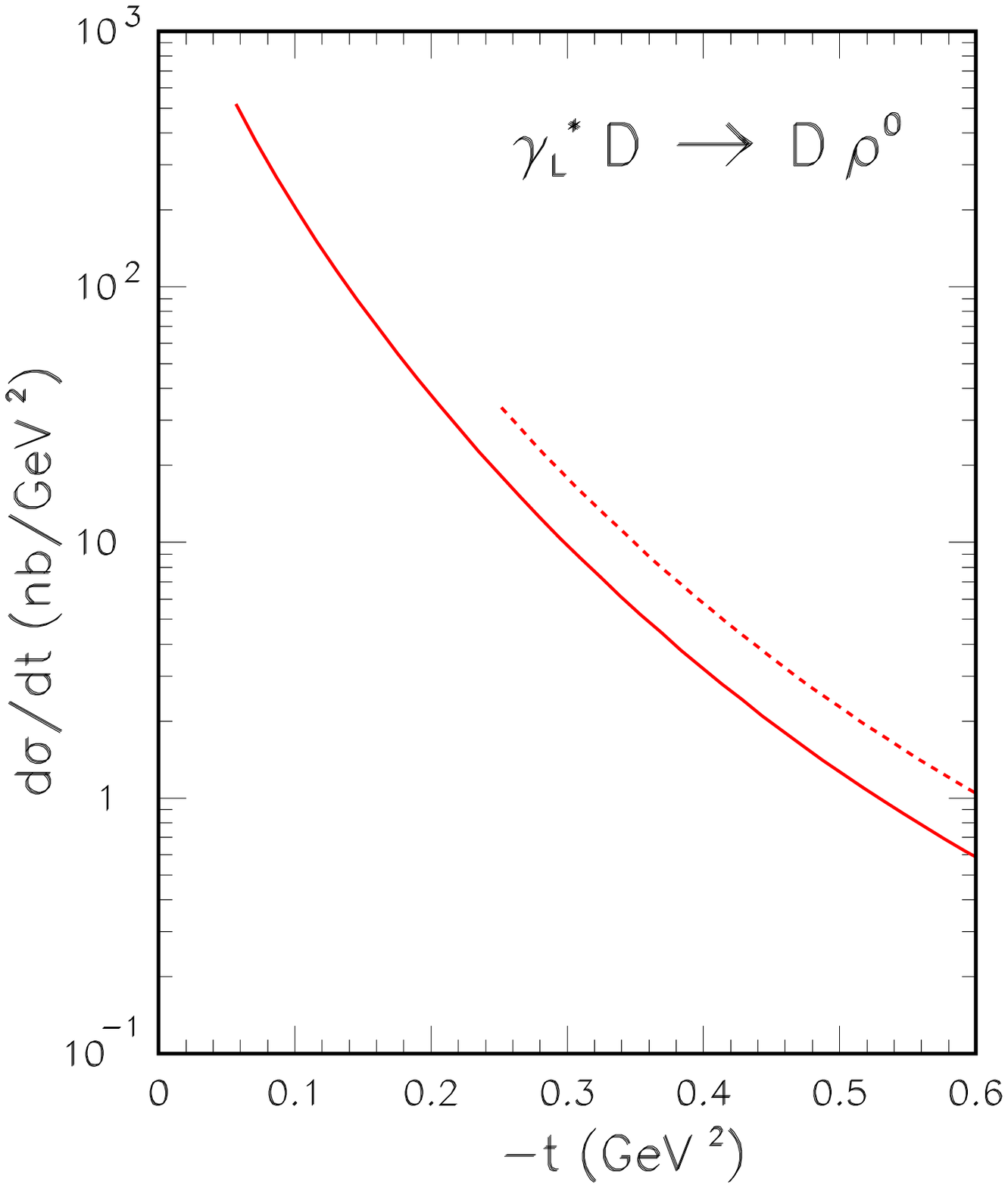} &
\includegraphics{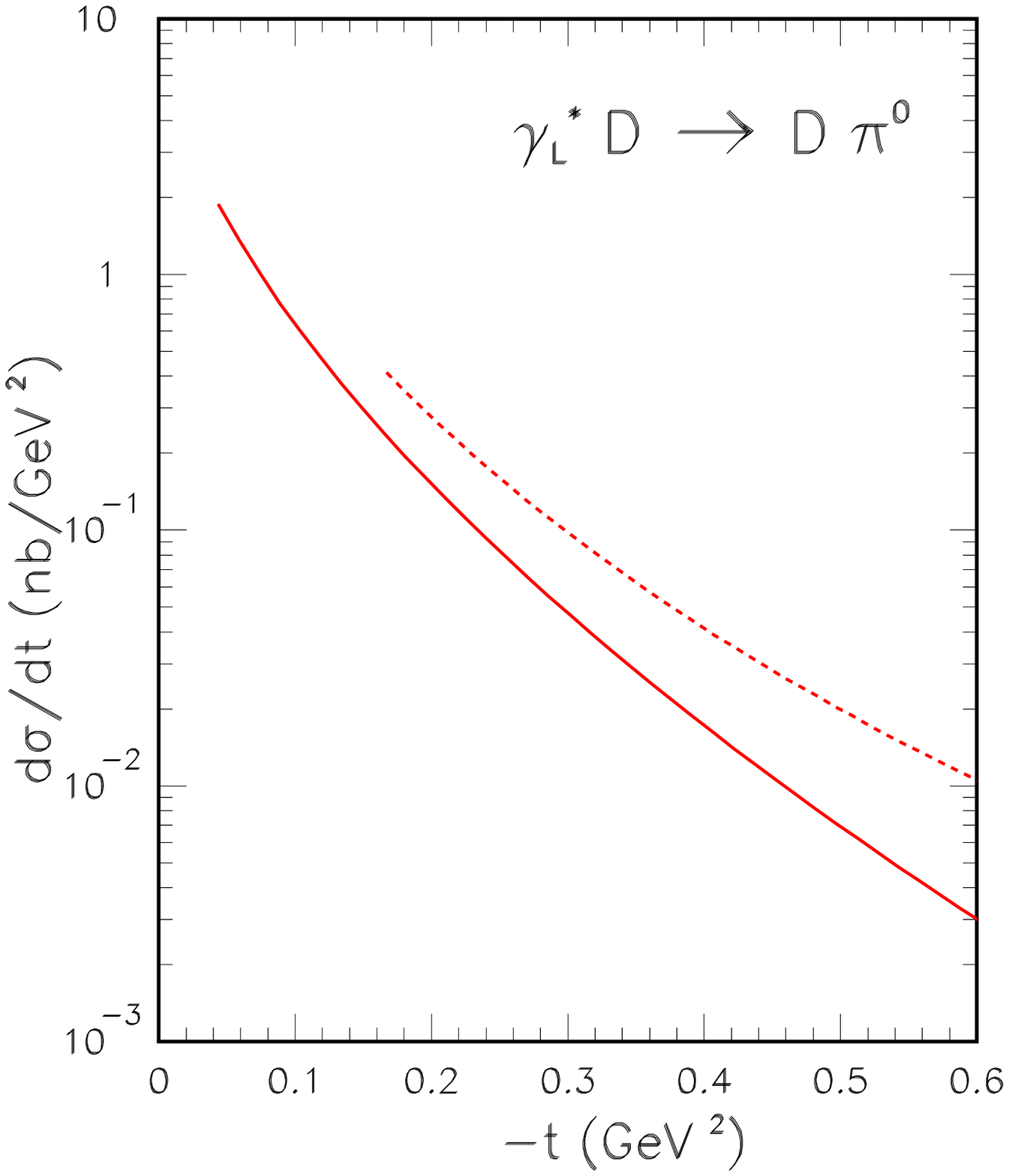}
\end{tabular}
}
\caption{$\rho^{0}$ (left) and $\pi^0$(right) virtual photoproduction cross
  sections at $\xbj=0.1$, $Q^2= 4$ GeV$^2$ (solid
 lines) and  $\xbj=0.2$, $Q^2= 3$ GeV$^2$ (dashed
 lines).}
\label{rhopi}
\end{figure}

\section{Conclusion}
We have demonstrated that deeply virtual Compton scattering and deep exclusive meson
 electroproduction on the deuteron is a feasible and promising field of study of the deuteron structure.
More theoretical work is obviously needed before one can draw definite conclusions from 
forthcoming data. In particular, some of the so-called higher twist terms are 
needed\cite{Anikin:2000em}
and should be estimated,
at least those coming from target mass effects\cite{BELITSKY01} that one may expect to 
be non-negligible. A first estimate of this effect has been given in the second
reference of Ref. 9.
The inverse process, photoproduction of a lepton pair, sometimes called
timelike Compton scattering is also feasible, and should give complementary
information as was shown in Ref.\cite{TCS} for the nucleon case. Double deeply 
virtual Compton scattering \cite{DDVCS}  may also be studied along the same lines. 
We expect that a deepened understanding of  the short distance structure of the deuteron will 
emerge from  these studies.

\vskip 0.5cm
We acknowledge useful discussions with M. Diehl, M. Gar{\c c}on and F. Sabatie. Special thanks to
 P.A.M. Guichon and M. Vanderhaeghen for their help with technical calculations.
 This work was supported by the EC--IHP Network ESOP, Contract HPRN-CT-2000-00130.

\renewcommand{\thesection}{\Alph{section}}
\setcounter{section}{0}

\section{Appendix}
\subsection{Light-cone Deuteron Wave Function}

We have chosen a covariant normalization for the one-particle states on the light cone:

\begin{eqnarray}
\langle p'^{\,+} , \vec{p}_\perp \,', \lambda ' | p^+, \vec{p}_\perp , 
\lambda  \rangle &=&
(2 \pi)^3 2 p^+ \delta( p'^{\,+} - p^{+}) \nonumber\\
&.&\delta^{(2)}( \vec{p}_\perp \,' -  \vec{p}_\perp ) \delta_{\lambda
\lambda '}
\end{eqnarray}

\noindent where $p^+$ is defined in terms of the ordinary vector components as $p^+ = 
\frac{1}{\sqrt{2}}(p^0 + p^3)$ and $\vec{p}_\perp$ corresponds to the components in 
the transverse direction. We have to evaluate matrix elements of a quark operator between 
deuteron states but eventually we have to deal with two quasi free nucleon states. The 
definition of the state of a deuteron with momentum $P$ and polarization $\lambda$ in terms
 of two-nucleon state is:

\begin{eqnarray}
&&|P^+ , \vec{P}_{\perp}, \lambda \rangle  =  \frac{1}{(16 \pi)^3} 
\sum_{\lambda_1, \lambda_2} \int \frac{d\xi_1}{\sqrt{\xi_1}} 
 \frac{d\xi_2}{\sqrt{\xi_2}}
\; \delta (1- \xi_1 -\xi_2 ) \nonumber \\
 & \cdot &  d\vec{p}_{1  _\perp}  d\vec{p}_{2  _\perp} \; \delta^{(2)}
  (\vec{P}_\perp - \vec{p}_{1  _\perp} - \vec{p}_{2  _\perp}) 
  \chi_\lambda (\xi_1, \vec{k}_{1  _\perp}, \lambda_1 ; \xi_2, \vec{k}_{2  _\perp},
   \lambda_2 ) \nonumber \\
& \cdot &  | p^+_1 , \vec{p}_{1  _\perp} , \lambda_1 ; p^+_2 ,
\vec{p}_{2 _\perp} , \lambda_2 \rangle 
\label{deuteronstate}
\end{eqnarray}

\noindent where $\xi_i = \frac{p^+_i}{P^+}$ and the transverse momenta $\vec{k}_{i  _\perp}$ are the transverse momenta of the nucleons in a frame where $\vec{P}_\perp =0 $ (hadron frame). We have also kept the notation $\vec{p}_{i _\perp}$ to make explicit the fact that the states are defined in an arbitrary frame but the wave function always refer to the hadron frame. The relationship between both set of coordinates is given by  

\begin{equation}
\vec{k}_{i  _\perp} = \vec{p}_{i  _\perp} - \xi_i  \vec{P}_\perp 
\end{equation} 

Let us define the measures that take part in the integrals as:

\begin{eqnarray}
[d\xi] & = & d\xi_1 d\xi_2\; \delta (1- \xi_1 -\xi_2 ) \\
 \left[ d\vec{p}_{\perp}\right] & = & d \vec{p}_{1  _\perp}  d\vec{p}_{2  _\perp} \; \delta^{(2)} (\vec{P}_\perp - \vec{p}_{1  _\perp} - \vec{p}_{2  _\perp})
\end{eqnarray}

We can also perform the integrals over the transverse momenta in the hadron frame with the measure:

\begin{equation}
 [ d\vec{k}_{\perp} ]  =  d \vec{k}_{1  _\perp}  d\vec{k}_{2  _\perp} \; \delta^{(2)} (\vec{k}_{1  _\perp} + \vec{k}_{2  _\perp})
\end{equation}

\noindent The wave function of the deuteron with polarization 
$\lambda$ is $\chi_\lambda (\xi_1, \vec{k}_{1  _\perp}, \lambda_1 ; \xi_2, 
\vec{k}_{2  _\perp}, \lambda_2 )$ with $\lambda_i$ being the polarization of the  nucleons. 
It is normalized according to

\begin{equation}
\sum_{\lambda_1, \lambda_2} \int \left[ d \xi \right] [ d\vec{k}_{\perp} ] | \chi_\lambda (\xi_1, \vec{k}_{1  _\perp}, \lambda_1 ; \xi_2, \vec{k}_{2  _\perp}, \lambda_2 )|^2 = 1 
\end{equation}

By taking advantage of the properties $\xi_1 \equiv \xi = 1 - \xi_2 $ and
 $\vec{k}_{1  _\perp} \equiv \vec{k}_{\perp} = -\vec{k}_{2  _\perp}$ we can further 
 simplify the notation in the wave function:

\begin{equation}
\chi_\lambda (\xi_1, \vec{k}_{1  _\perp}, \lambda_1 ; \xi_2, \vec{k}_{2  _\perp}, \lambda_2 ) \equiv 
\chi_\lambda (\xi, \vec{k}_{\perp};  \lambda_1, \lambda_2 )
\end{equation}

A last remark concerns the connection between the light-cone wave function of the deuteron 
and the ordinary (instant-form) relativistic wave function obtained with different phenomenological 
potentials. Whereas the first one is expressed in terms of light-cone coordinates, the latter one is a 
function of the ordinary three-vectors $\vec{k}_i$ and fulfills, in general, a Schr{\"o}dinger type 
equation. It can be shown \cite{BAKKER79} that, if we define the longitudinal momentum from 
the light-cone coordinates as:

\begin{equation}
{k_z}\equiv M_0 (\xi - 1/2)
\end{equation} 

\noindent where $M_0$ is the free mass operator:

\begin{equation}
M_0^2 = \frac{m^2 + \vec{k}_{\perp} \,^2}{\xi (1-\xi)}
\end{equation}

\noindent then the eigenvalue  equation fulfilled by $\chi_\lambda (\xi, \vec{k}_{\perp};
  \lambda_1, \lambda_2 )$ can be interpreted as a Schr{\"o}dinger equation and therefore, it can
   be related to the wave function obtained from phenomenological potentials in the instant-form 
   formalism. More explicitely:

\begin{eqnarray}
&&\chi_\lambda (\xi, \vec{k}_{\perp};  \lambda_1, \lambda_2 )  =  \sum_{\mu_1, \mu_2} \left[ \frac{M_0}{4 \xi (1 -
\xi)}\right]^{1/2} \langle \lambda_1 |R_M^\dag (\xi, \vec{k}_{\perp})| \mu_1 \rangle \nonumber \\
& \cdot & \langle \lambda_2 |R_M^\dag (1- \xi, -\vec{k}_{\perp})| \mu_2 \rangle \; \chi_\lambda^c ( \vec{k} ;  \mu_1, \mu_2 )
\end{eqnarray}

The global factor in the r.h.s. of the equation above is just the Jacobian of the transformation from 
the variables  $\{ \xi, \vec{k}_{\perp}\}$ to $\{ \vec{k}\}$. We also have the matrix elements 
of the Melosh rotation, which relate the spin in the light-front with the spin in the instant-form of the
 dynamics. Finally, the (canonical) deuteron wave 
 function $\chi_\lambda^c ( \vec{k} ;  \mu_1, \mu_2 )$ is written as \cite{CHUNG88}:

\begin{eqnarray}
\chi_\lambda^c ( \vec{k} ;  \mu_1, \mu_2 ) &=& (16 \pi^3)^{1/2} 
\sum_{L,m_L,m_s} (\frac{1}{2}\, \frac{1}{2} \, 1 \, | \,
\mu_1 \, \mu_2 \, m_s \,)\nonumber\\
&& (L \, 1 \, 1 \, | m_L \, m_s \, \lambda ) Y_{L,m_L}(\hat{k}) u_L(k) 
\end{eqnarray}

\subsection{Helicity amplitudes and GPDs}

We give here the kinematical coefficients that relate the helicity amplitudes evaluated with the 
light-cone helicity vectors and the generalized parton distributions. To be consistent with the choice 
of a right-handed set of polarization vectors made in \cite{BERGER01}, these expressions should 
be used only if $\Delta_x < 0$ in the explicity evaluation of $V_{\lambda',\lambda}$ or
 $A_{\lambda',\lambda}$.

By applying parity properties we can reduce the number of independent helicity amplitudes 
$V_{\lambda',\lambda}$ down to five (that we have chosen to be $V_{++}$, $V_{00}$, 
$V_{-+}$, $V_{0+}$, $V_{+0}$) and to four in the pseudovector case $A_{\lambda',\lambda}$ 
(we keep $A_{++}$, $A_{-+}$, $A_{0+}$, $A_{+0}$). We recall these parity and time reversal 
properties:

\begin{eqnarray}
V_{\lambda ' \lambda} & = & (-1)^{\lambda ' - \lambda} V_{-\lambda ' -\lambda} \\
A_{\lambda ' \lambda} & = & - (-1)^{\lambda ' - \lambda} A_{-\lambda ' -\lambda} \\
 & & \;  \nonumber \\
V_{\lambda ' \lambda}(\xi) & = & (-1)^{\lambda ' - \lambda} V_{\lambda \lambda '}(-\xi) \\
A_{\lambda ' \lambda}(\xi) & = & (-1)^{\lambda ' - \lambda} A_{\lambda \lambda '}(-\xi)\\
\end{eqnarray}

We denote:

\begin{eqnarray}
H_i & = & \sum_{\lambda', \lambda} c_i^{\lambda ' \lambda} V_{\lambda ' \lambda} \\
\tilde{H}_i & = & \sum_{\lambda', \lambda} \tilde{c}_i^{\lambda ' \lambda} A_{\lambda ' \lambda}
\end{eqnarray}

\noindent where the sum covers only the helicity amplitudes that we have chosen as independent ones and the non-vanishing coefficients are: 

\begin{eqnarray}
c_1^{++} & = & \frac{1}{3 (1-\xi^2)^2}(3 \xi^4 - 7 \xi^2 - 2 D (1-\xi^2)+2) \nonumber \\
c_1^{00} & = & \frac{1}{3 (1-\xi^2)} \nonumber \\
c_1^{-+} & = & - \frac{1}{3 D (1-\xi^2)^3}(2 \xi^2 +D (3 \xi^6 -10 \xi^4 + 9 \xi^2 -2)) \nonumber \\
c_1^{0+} & = & \frac{2}{3(1-\xi^2)^2} \sqrt{\frac{1+\xi}{2 D(1-\xi)}}( D (1-\xi^2)+\xi) \nonumber \\
c_1^{+0} & = & - c_1^{0+}(\xi \rightarrow - \xi)  \\
 & & \;  \nonumber \\
c_2^{++} & = & \frac{2}{1-\xi^2} \nonumber \\
c_2^{-+} & = & \frac{2\xi^2}{D(1-\xi^2)} \nonumber \\
c_2^{0+} & = & - \frac{1}{1-\xi} \sqrt{\frac{1+\xi}{2 D(1-\xi)}} \nonumber \\
c_2^{+0} & = & - c_2^{0+}(\xi \rightarrow - \xi) \\
& & \;  \nonumber \\
c_3^{-+} & = & - \frac{1}{ D} \nonumber \\
& & \;  \nonumber \\
c_4^{++} & = & - \frac{2\xi}{1-\xi^2} \nonumber \\
c_4^{-+} & = & - \frac{2\xi}{D(1-\xi^2)^2} \nonumber \\
c_4^{0+} & = &  \frac{1}{1-\xi} \sqrt{\frac{1+\xi}{2 D(1-\xi)}} \nonumber \\
c_4^{+0} & = &  c_4^{0+}(\xi \rightarrow - \xi) \\
& & \;  \nonumber \\
c_5^{++} & = & - \frac{1}{ (1-\xi^2)^2}(\xi^2 + 2 D (1-\xi^2)+1) \nonumber \\
c_5^{00} & = & \frac{1}{ (1-\xi^2)} \nonumber \\
c_5^{-+} & = & - \frac{1}{ D (1-\xi^2)^3}(2 \xi^2 +D(1-\xi^4)) \nonumber \\
c_5^{0+} & = & \frac{2}{(1-\xi^2)^2} \sqrt{\frac{1+\xi}{2 D(1-\xi)}}( D (1-\xi^2)+\xi) \nonumber \\
c_5^{+0} & = & - c_5^{0+}(\xi \rightarrow - \xi)  
\end{eqnarray}

 In the pseudovector case they are :

\begin{eqnarray}
\tilde{c}_1^{++} & = & \frac{1}{ ( 1 + D (1-\xi^2))} \nonumber \\
\tilde{c}_1^{-+} & = & \frac{D (1-\xi^2)}{ \xi ( 1 + D (1-\xi^2))} \nonumber \\
\tilde{c}_1^{0+} & = & \frac{(1+\xi)\sqrt{2 D (1-\xi^2)}}{2 \xi ( 1 + D (1-\xi^2))} \nonumber \\
\tilde{c}_1^{+0} & = & - \tilde{c}_1^{0+}(\xi \rightarrow - \xi)  \\
 & & \;  \nonumber \\
\tilde{c}_2^{++} & = & \frac{1}{4 ( 1 + D (1-\xi^2))} \nonumber \\
\tilde{c}_2^{-+} & = & \frac{\xi^2 - D (1-\xi^2)^2}{4 D \xi (1-\xi^2) ( 1 + D (1-\xi^2))} \nonumber \\
\tilde{c}_2^{0+} & = & -\frac{(1+\xi)}{4 \xi \sqrt{2 D (1-\xi^2)} ( 1 + D (1-\xi^2))} \nonumber \\
\tilde{c}_2^{+0} & = & - \tilde{c}_2^{0+}(\xi \rightarrow - \xi)  \\
 & & \;  \nonumber \\
\tilde{c}_3^{++} & = & - \frac{\xi}{4 ( 1 + D (1-\xi^2))} \nonumber \\
\tilde{c}_3^{-+} & = & - \frac{1}{4 D (1-\xi^2) ( 1 + D (1-\xi^2))} \nonumber \\
\tilde{c}_3^{0+} & = & \frac{(1+\xi)}{4  \sqrt{2 D (1-\xi^2)} ( 1 + D (1-\xi^2))} \nonumber \\
\tilde{c}_3^{+0} & = & \tilde{c}_3^{0+}(\xi \rightarrow - \xi)  \\
 & & \;  \nonumber \\
\tilde{c}_4^{++} & = & - \frac{1}{ 1-\xi^2} \nonumber \\
\tilde{c}_4^{-+} & = & - \frac{\xi}{ D (1-\xi^2)^2} \nonumber \\
\tilde{c}_4^{0+} & = & \frac{1}{(1-\xi) \sqrt{2 D (1-\xi^2)}} \nonumber \\
\tilde{c}_4^{+0} & = & -\tilde{c}_4^{0+}(\xi \rightarrow - \xi)  
\end{eqnarray}

In the forward limit ($\xi \rightarrow 0, t \rightarrow 0$) we get the simplified expressions:

\begin{eqnarray}
H_1 (x,0,0)& = & \frac{1}{3}(2 V_{++}+ V_{00}) \\
H_5 (x,0,0)& = & (- V_{++} + V_{00}) \\
\tilde{H}_1 (x,0,0)& = &  A_{++}
\end{eqnarray}

The integral over $H_1 (x,0,0)$ is the parton number, which imposes a serious check on the 
contruction of the helicity amplitudes in the forward limit.
\subsection{Parameterization of the nucleon matrix elements}

In a symmetric frame, where $\bar{P}^\mu=(p^\mu + p'^\mu)/2$ has no transverse 
momentum we take the following parameterization:

\begin{eqnarray}
&&\int \frac{d\kappa}{2 \pi} e^{i\kappa x} \; \langle p', \lambda '
 | \bar{\psi}_q(-\frac{\kappa}{2}) \gamma^+
\psi_q(\frac{\kappa}{2})| p , \lambda \rangle  =   \\
&&\bar{u}(p',\lambda ') \left[ \gamma^+ H^q(x,\xi,t) + i \frac{\sigma^{+ \alpha} 
\Delta_\alpha}{2 m} E^q(x,\xi,t)\right]
u(p,\lambda) \nonumber \\
&&\int \frac{d\kappa}{2 \pi} e^{i\kappa x} \; \langle p', \lambda '
 | \bar{\psi}_q(-\frac{\kappa}{2}) \gamma^+ \gamma_5
\psi_q(\frac{\kappa}{2})| p , \lambda \rangle  =   \\
&&\bar{u}(p',\lambda ') \left[ \gamma^+ \gamma_5  \tilde{H}^q(x,\xi,t) + 
\frac{\gamma_5 \Delta^+}{2 m}
\tilde{E}^q(x,\xi,t)\right] u(p,\lambda)  \nonumber
\end{eqnarray}

\noindent where the integration path must be understood along the '-' direction. By using 
light-cone helicity states,
which is close to the usual helicity in frames where the particle moves fast to the right and 
allows to  get compact and
elegant expressions, we have to insert in the previous equations the following results:
\begin{eqnarray}
&&\bar{u}(p',\lambda ') \gamma^+ u(p,\lambda)  =   2 \bar{p}^+ \sqrt{1-\xi^2}
 \delta_{\lambda \lambda '} \nonumber\\
&&\bar{u}(p',\lambda ') \gamma^+ \gamma_5 u(p,\lambda)  =   2 \lambda 2 \bar{p}^+ \sqrt{1-\xi^2} \delta_{\lambda
\lambda '} \nonumber\\
&&\bar{u}(p',\lambda ') i \frac{\sigma^{+ \alpha} \Delta_\alpha}{2 m} 
u(p,\lambda)  =   2 \bar{p}^+ \\
&&~~~~ \cdot \left(  -
\frac{\xi^2}{\sqrt{1-\xi^2}} \, \delta_{\lambda \lambda '} +  
\frac{\sqrt{t_0-t}}{2 m} \eta_\lambda \delta_{\lambda,-
\lambda '} \right) \nonumber\\
&&\bar{u}(p',\lambda ') \frac{\Delta^+}{2 m } \gamma_5 u(p,\lambda)  =  
 2 \lambda 2 \bar{p}^+ \nonumber\\
&&~~~~  \cdot \left(  -
\frac{\xi^2}{\sqrt{1-\xi^2}} \, \delta_{\lambda \lambda '} + 
 \xi \frac{\sqrt{t_0-t}}{2 m} \eta_\lambda
\delta_{\lambda,-\lambda '} \right) \nonumber
\end{eqnarray}

\noindent where 
\begin{equation}
\eta_\lambda = \frac{2 \lambda \Delta_x - i \Delta_y}{|\mathbf{\Delta}_\perp|}
\end{equation}
and $\Delta^\mu = p'^\mu - p^\mu$.

\end{document}